\documentclass[draft]{article}

\usepackage{epsf}
\usepackage{amsfonts}
\usepackage{amssymb}
\usepackage{amsmath}
\usepackage[pagebackref]{hyperref} 
\usepackage{pstricks}
\usepackage{fullpage}
\def\etal{{\it et.al.}\ }

\def\names{\doteq}

\def\eqn#1{Eqn.(\ref{#1})}
\def\fig#1{Fig. (\ref{#1})}
\def\figdir{.}

\newcounter{thedefn}
\newtheorem{mydefinition}[thedefn]{Definition}
\newcounter{theprin}
\newtheorem{principle}[theprin]{Principle}

\begin{document}
\title{What is an OS?}
\author{Abhijat Vichare,
\\Computational Research Laboratories Ltd.\\Damle Path, Off Law
 college Rd., Pune 411004, INDIA.\\Email: \texttt{abhijatv@gmail.com}}
\maketitle

\begin{abstract}
  While the engineering of operating systems is well understood, their
  formal structure and  properties are not.  The latter  needs a clear
  definition  of the purpose  of an  OS and  an identification  of the
  core.  In  this paper I offer  definitions of the  OS, processes and
  files, and present a few useful principles.  The principles allow us
  to  identify  work  like  closure and  continuation  algorithms,  in
  programming  languages  that is  useful  for  the  OS problem.   The
  definitions and  principles should  yield a symbolic  framework that
  encompasses practice.  Towards that end I specialise the definitions
  to describe conventional OSes and identify the core operations for a
  single computer  OS that  can be used  to express  their algorithms.
  The  assumptions underlying  the algorithms  offer the  design space
  framework.  The paging  and segmentation algorithms for conventional
  OSes  are  extracted from  the  framework  as  a check.   Among  the
  insights  the  emerge is  that  an OS  is  a  constructive proof  of
  equivalence  between  models   of  computation.   Clear  and  useful
  definitions  and  principles are  the  first  step  towards a  fully
  quantitative structure of an OS.
\end{abstract}


\section{Introduction}
\label{sec:introduction}

Operating systems (OS, plural:  OSes) are well understood in practice.
Most of  the work  in OS  has been technological,  and as  Yates \etal
describe  succinctly  in  \cite{Yates99i/oautomaton}:  ``very  systems
oriented and  results driven''.  Significant effort has  been spent on
the ``How to  have an OS?'', some  on the ``Why to have  an OS?'', and
almost none  on the ``What  is an OS?''.   The ``How'' of an  OS often
concerns itself  with resource management algorithms.   The ``Why'' is
usually answered in  terms of ``end user convenience'',  or better, an
``abstract  machine''.   It  is  not  clear if  there  are  any  other
fundamental reasons.   For instance, if ``Why  manage resources?''  is
answered as:  ``To have an  abstract machine'', then the  natural next
question  is:  ``What  is   the  desired  abstract  machine?''.   Such
questions are  often only  intuitively clear, and  need to  be clearly
asked and answered if we are to seek formal, mathematical descriptions
of OS.   Characterizing an OS, i.e.  answering  the ``What'' question,
subsumes  the  answers to  both  the  above  questions.  Bergstra  and
Middelburg  have  recently  pointed   out  this  issue  explicitly  in
\cite{DBLP:journals/corr/abs-1006-0813}: the absence of the definition
of  what an OS  is.  They  also add  another issue:  the absence  of a
formal  view of  OS.  In  fact,  Middelburg documents  his efforts  of
surveying   the   literature   precisely   for  these   questions   in
\cite{DBLP:journals/corr/abs-1003-5525}, and  concludes that (a) there
is  ``only   one  more   or  less  abstract   model  by   Yates  \etal
\cite{Yates99i/oautomaton}'',  (b)  there ``does  not  exist a  theory
based on that model'', and (c) that the OS community ``has paid little
attention  to  clarifying  what  an  OS  is  and  giving  motives  for
introducing OSes''.  Attempts at a high level view of OSes also occur.
For         example,         Ramamritham         and         Stankovic
\cite{Ramamritham94schedulingalgorithms}    discuss   four   paradigms
underlying  scheduling   in  real  time  OSes:   static  table  driven
scheduling,    static   priority   preemptive    scheduling,   dynamic
planning-based  scheduling, and dynamic  best effort  scheduling.  The
broad survey in section  \ref{sec:lit:review} also shows that although
the  OS  problem  has  been  looked  at from  a  number  of  different
perspectives there  is no significant effort to  crystallize the ideas
clearly into a  formal structure.  Developing an answer  to the ``What
is  an OS?''  question  and exploring  the quality  of its  connect to
known practice would form the initial step toward this goal.

Identification  of essential abstractions  is necessary  to supplement
systems oriented work to deal with its rapidly growing complexity.  At
its core,  an OS is a  process that bridges two  execution levels, low
and high.  If an OS provides an execution environment to its programs,
then its  behavior at  the two  levels ought to  be described  by some
theory of processes.  Several  attempts to formally describe processes
have     been,      and     are     being      made     (e.g.      see
\cite{Milner:1982:CCS:539036},         \cite{Brookes:1984:TCS:828.833},
\cite{bergstra:pa},  \cite{VanHorn:2010:AAM:1932681.1863553}, and more
recently   \cite{Nielson:2012:FLP:2071389.2071392})  with  concurrency
issues  as the  main concern  today.  We  will refer  to  these varied
approaches as ``process algebras''  in this work.  The process algebra
for  an   OS  would  need  a  clear   identification  and,  preferably
constructive, definition  of core concepts and  principles guiding the
formalization effort.  The definitions  must concretize the two levels
and along with  the principles, help in devising  algorithms to bridge
the  gap between  the two  execution levels.   This paper  presents an
effort towards  the ``What'', and  offers two main  contributions: (a)
definitions of an OS and a set of principles that could be a part of a
theory  of OS,  (b) using  them to  devise the  formal  algorithms for
current OSes  in practice that  systematically bridge the  gap between
the two defined levels.  The  unambiguous definition of the two levels
should hopefully enable constructing a  process algebra for OS that is
similar  to the  algebras  in programming  languages.   After all  the
semantics of the expressed and the executed ought to match, and should
be  a part of  a theory  of OS  that complements  a theory  of program
expression.

Further  work  requires   explicative  definitions  built  upon  known
concepts and principles that capture the problem invariants.  They can
be used to build a framework that organizes knowledge and offer useful
insights.   The goal  of  this  paper is  to  examine definitions  and
principles  for OS  over  single processor  machines.  The  simplified
problem reduces system complexity, and offers a well known practice as
the  base to  verify against.   To verify  I use  the  definitions and
principles to define the  conventional OS in practice and symbolically
express  the  main  algorithms.   I identify  the  essential  abstract
operations for a  typical OS that can be used  to formally express its
algorithms.   I show  that the  main algorithms  for OSes  over single
processor systems can emerge from these expressions.  The technique is
illustrated for one algorithm.   The proposed framework can manage the
complexity  of OS for  single processor  systems, and  can potentially
also  separate  the formal  issues  like  correctness from  technology
issues   like  performance.   Correctness   issues  need   the  formal
properties of the  abstract operations for the OS  problem, and do not
form the  focus of  the current work.   The focus on  single processor
OSes is a limitation of this  work.  The essential idea in the work is
that a common execution model is the central concept that brings OSes,
programming  languages  and  systems  software  together.   While  the
individual  trees  are no  doubt  important,  a  shift in  perspective
creates an equally important garden.

To  reach the  goal, this  paper considers  the OS  over  a simplified
system, and discards some important issues.  The underlying model is a
single CPU, single primary memory  system with a clock interrupt.  Any
practical OS spends a  substantial effort dealing with I/O.  Howsoever
significant in practice, in this work we see I/O as source or sinks of
(byte)  streams.   Practical OSes  today  support  IPC  and deal  with
associated problems  like critical sections and  deadlocks.  These are
issues of  the execution model and  process algebras are  used to deal
with them.   We leave such  issues to process algebras;  e.g.  process
algebras may  capture deadlocks as an  action that is the  zero of the
sequencing operation  and the  identity of the  alternation operation.
However, we do aim to examine a way to bridge the execution levels and
identify any structure within.  Such  a base would be needed to ensure
matching the  program expression and the  program execution semantics.
Current  challenges in  multicore,  cluster, and  true concurrency  in
distributed systems are also not considered as a part of the work.

\subsection{Organization of the work}
\label{sec:organisation}

I start by  offering definitions for an OS, a process  and a file, and
discussing  them.   The motivations  from  programming languages  that
yield the  precise description of an  OS also offer the  base for some
useful  principles.   The  proposed  definitions and  principles  must
capture conventional  OSes.  Hence I first  specialize the definitions
to  express conventional  OSes,  motivate the  need  and identify  the
abstract operations  needed to satisfy the definition,  and employ the
principles to  devise the abstract  algorithms for the usual  OSes.  A
specific algorithm in use is instantiated from the abstract algorithms
to  connect with practice,  and can  be found  in the  appendix.  This
demonstrates  that the  current work  can indeed  capture conventional
OSes, and hence can claim that the proposed definitions and principles
have  potential towards  developing a  theory of  OS.

\section{What is an OS?}
\label{sec:what:is:an:os}

Programming  languages as  mechanisms  of program  expression have  an
execution model  that executes  the expressed programs.   OSes execute
programs and  need an execution model.  Execution  models are supplied
by  formal  models of  computation,  and  are  the connecting  concept
between program  expression and program execution.  They  serve as the
base for defining  OSes.  An OS can be defined  either using models of
computation or using programming languages.

\subsection{Definition of an OS}
\label{sec:os:defn}
\label{sec:defns}

\begin{mydefinition}
  \label{basic:os:def:1}
  An \emph{Operating System} is an universal machine that implements a
  Turing complete model of computation, called the high level machine,
  in  terms of  another Turing  complete model,  called the  low level
  machine.
\end{mydefinition}

Three  of  the  many  equivalent  views of  computation  are  language
recognition  exemplified by  the Turing  machines model,  the function
computation view exemplified by  the $\lambda$ calculus model, and the
symbol  transduction view  exemplified by  the Markov  algorithms view
(e.g. see  \cite{Taylor:1998:MCF:275566}).  An  OS would be  a program
that  would implement  a  model of  computation  given another  model.
Given a Turing  machine based system, an OS  would provide a $\lambda$
calculus based one,  or given a Markov algorithms  based system, an OS
would provide a Turing machines based  system, and so on.  One may not
have an  OS between  a low level  push down  machine and a  high level
Turing machine  since the  computational power of  the low  level push
down machine  is less  than a Turing  machine.  In practice,  the high
level machine  is empirically and arbitrarily defined,  and is loosely
described as ``being  useful to the end user''.   The definition above
offers a way to overcome this deficiency as we shall see later in this
work.  An OS is usually described  as a program that provides a useful
abstract machine  over the underlying  raw machine.  In that  sense it
bridges  the ``gap'' between  a low  level Turing  complete imperative
machine,  and  a  high  level   one.   The  gap  is  intuitively  well
understood.

An alternate definition of an OS is:
\begin{mydefinition}
  {An} \emph{Operating  System} is any process that  can interpret one
  or more of its procedures that are defined at its run time.
  \label{def:os:constructive}
\end{mydefinition}

As  an interpreter  of procedures  defined at  its runtime,  an  OS is
essentially  an   universal  machine.   It   accepts  descriptions  of
procedures and  interprets them.  The  descriptions may be  defined at
the runtime of  the OS and have some intended  semantics.  Given a set
of  procedures,  a universal  machine  must  interpret each  procedure
without violating its semantics.  An  execution sequence of the set of
procedures is a permutation of the given set subject to the constraint
that the semantics of each are  not affected by the OS.  We will refer
to  this  constraint  as  \emph{orthogonal execution  of  procedures}.
Every possible execution sequence  that allows orthogonal execution of
procedures  is  admissible,  and  must  be  supported.   An  execution
sequence also corresponds to the  formal computation by the high level
abstract machine.   As a universal machine, the  formal computation by
an OS is  completely determined by its own  computation and the formal
computation by  each of the  procedures.  An OS  ensures orthogonality
between  procedure execution and  its own  execution by  providing per
procedure  contexts for  interpretation,  and is  said  to define  the
execution environment for a procedure.

The   obvious    operations   on   the   set    of   procedures   are,
\emph{add-procedure}      i.e.       program      invocation,      and
\emph{delete-procedure}  i.e.  program  exit.  Between  the invocation
and  exit, an  OS must  support all  admissible execution  patterns by
defining  some  primitives that  can  be  used  to specify  a  desired
pattern.  The primitives  allow an OS to freely  change the procedures
between  active (i.e.   under interpretation)  and passive  (i.e.  not
being  interpreted)  states.  These  procedures  must  be first  class
objects in an OS for such primitives.  To transit from an active state
to a passive state, an OS computes the closures and continuations.  To
transit from a passive state to  an active state, an OS restores them.
First classness is required to  evolve an ongoing execution pattern of
the set of programs.

The execution  patterns to be  supported can be obtained  by examining
the  implicit   assumptions  behind  the  execution   models  used  by
programming languages.   For instance, an  activation stack discipline
of  execution  is  related  to  the  assumption that  a  caller  of  a
subprogram is  dormant once it  initiates the subprogram.  It  is also
related to the assumption that  a caller activates another callee only
after   the  current  callee   returns.   Relaxing   such  assumptions
introduces  the more expressive  coroutines, and  is identical  to the
simultaneous  execution of processes  in a  cooperatively multitasking
OS.  The execution environment that an OS has to support can therefore
be  obtained  by  relaxing   such  implicit  assumptions  for  program
expression.  In general, relaxing such assumptions amounts to creating
the weakest, i.e. the most general, form of procedure activation.  For
an OS, the ability to run  the procedures in all possible ways is thus
a matter of supporting  an execution environment commensurate with the
expressiveness,  and  operational efficiency  is  a by-product.   This
departs from  the central managerial role that  practice often accords
an OS, but does not ignore it.

\subsubsection{Process and File}
\label{sec:process:and:file}

An operating  system deals with  two entities: a \emph{process}  and a
\emph{file}.  A process is often defined as ``an instance of a program
in   execution'';   see  \cite{Denning:1971:TGC:356593.356595}.    The
definition  is   esoteric  since  a  process   is  understood  \emph{a
  posteriori}, i.e. after one has  some experience with the ideas of a
``program  instance''  and its  ``execution''.   Dennis  and van  Horn
\cite{Dennis:1966:PSM:365230.365252}  define a  process in  an \emph{a
  priori}, but  more abstract,  way as: ``A  process is  that abstract
entity  which moves  through the  instructions of  a procedure  as the
procedure is executed by a  processor.''.  However, it is not as clear
an operational view of a  process.  To add an operational component to
this  view we  use elementary  proof theory  (see Gries  and Schneider
\cite{Gries:1993:LAD:161182}, for example), since it precisely defines
interpretation.  Intuitively, interpretation is the ability to extract
the value  bound to a  given symbol.  As  an universal machine,  an OS
keeps  interpreting the  symbols  of its  procedures.   A process  is,
therefore, the formal computation by the high level machine when a set
of symbols of  a procedure are presented to  the interpreter.  The set
of symbols are represented as  natural numbers and will be referred to
as  \emph{procedures}.   Procedures  are  the final  goal  of  program
expression  while a  process is  specific to  program  execution.  The
conventional  definition of  a  process does  not sharply  distinguish
between the  final goal of  program expression and  program execution.
The overall system state is composed of the OS state and the states of
the set of  procedures it deals with.  A formal theory  of an OS would
prescribe  rules  for this  composition,  e.g.   the orthogonality  of
execution between an OS and  each of its procedures.  Process algebras
indeed  do  attempt  such  descriptions.  The  sequence  operation  is
associative  but not  commutative in  process algebra  for interacting
processes,  but  can be  commutative  for  non-interacting ones.   The
notion of a  process as formal computation is  also invariant over all
the  computer organization  and  computation organization  strategies,
whether single processor machines or distributed systems.

\begin{mydefinition}
  \label{def:process}
  A  \emph{process} is  the formal  computation by  a  formal machine,
  e.g. a Turing machine.
\end{mydefinition}

A file is  often defined simply as a ``collection  of bytes'', or more
precisely as an ``uninterpreted  collection of natural numbers''.  This
view is general  enough in that any data object  that is operated upon
by  a  process  is   a  \emph{file},  or  \emph{generalized  file}  to
distinguish from the conventional file.  It includes the usual notions
of variables, data structures and conventional files.

\begin{mydefinition}
  \label{def:file}
  A \emph{file} is a collection of one or more natural numbers.
\end{mydefinition}

A process  causes synchronized changes in  the state of  the system as
its interpretation progresses through the procedure in time, and is an
``active''        entity       (\cite{Denning:1971:TGC:356593.356595},
\cite{Holt:1972:DPC:850614.850627}).  A  file cannot ever  cause state
changes, and  is said to be  a ``passive'' entity.  In  essence, an OS
deals  primarily with  a set  of natural  numbers that  can  either be
active or passive.  Constructively, a collection of natural numbers is
a file if it  is passive and process if it is  active.  An OS requires
the first  classness for ``load''/``store'' operations as  a file, and
``schedule''  operations  as  a  process.   Procedures  correspond  to
operators and files correspond to operands in an application $\lambda$
term.

\subsection{Useful principles}
\label{sec:useful:principles}

Principles capture invariants of a problem that can be used to develop
a structure of  the solution.  The questions that  help capturing them
are: (a) What invariants can be induced from the definitions?, (b) How
can we identify the steps required by an OS as an universal machine to
bridge two  models of computation?, and  (c) Given the  steps what can
help their  correct arrangement?  This section looks  at the arguments
from programming languages to answer these questions, and utilize them
to obtain some general principles that  could be used in OS work.  The
principles are  employed to current  OS practice, by first  drawing an
analogy to the expression levels in programming languages.

\subsubsection{First classness}
\label{sec:first:classness}

The  definitions  suggest  first   classness  as  one  such  invariant
requirement  for the  OS  problem.  First  classness  is necessary  to
evolve  execution patterns,  and operates  over generalized  files.  A
file may  become a procedure if  it is scheduled to  consume CPU time,
and is a member of a set  used to decide or select for scheduling.  It
is  a parameter  to decision  functions that  fix the  schedule,  or a
return  value of  selection functions  that pick  a file  to schedule.
This general view of a file  can be used to express a principle useful
for OS, which we call ``principle of first classness''.
\begin{principle}
  \label{first:classness:principle}
  Generalized files in an OS must be first class objects.
\end{principle}

\subsubsection{OS and Programming Languages}
\label{sec:os:and:prog:lang}

While programming languages concern  themselves with the techniques of
expressing  programs,   operating  systems  concern   themselves  with
creating the  machines for their  execution.  However, an  OS requires
more  than a  common execution  model between  program  expression and
execution.   It   must  also  support  all   the  execution  sequences
consistent  with   orthogonal  execution.   A   language  defines  the
execution model for expression  via its interpreter.  A C interpreter,
for instance, has a stack  based execution model, provides a procedure
call and explicit memory management as a part of the expression model,
and offers little support to compute closures.  An OS requires closure
computation  for store-restore  operations, called  context switching.
If  expressed in  C, then  the OS  must implement  closure computation
since the C system has little  support for it.  An OS must incorporate
algorithms  to realize  the execution  patterns required  by  the high
level machine and not supported by the expression language.

As a  program, an  OS is  expressed using the  execution model  of the
language of expression.   It may choose to execute  over the execution
model of the  low level machine, or it may  execute over the execution
model  of  the  language  of  expression.  In  the  latter  case,  the
execution model  of the language  of expression is assumed  to subsume
the  low  level  machine  completely.   This is  not  always  true  in
practice,  and is a  consequence of  the trade off  between portability
across machines and specificness to  a given machine.  Thus, there are
three models of  execution at work: the low level  model over which an
OS may execute,  the high level target model that  an OS realizes, and
the model of expression used to express the OS.

\begin{figure}[t]
  \centering
  \epsfxsize=.5\textwidth
  \epsffile{\figdir/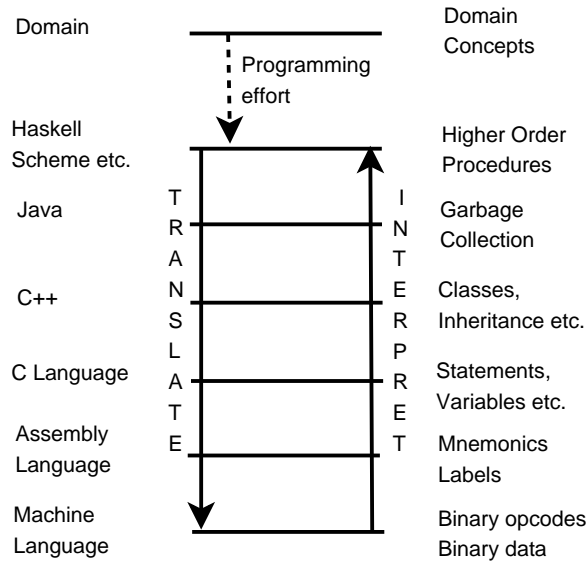}
  \caption[Abstraction    of   expression   levels    in   programming
  languages]{Abstraction of expression  levels in programming language
    design.  Programming effort is required to express domain problems
    at the chosen high level of abstraction.  The figure shows typical
    lowest and highest levels.}
  \label{fig:prog:lang:abstractions}
\end{figure}

An intuitive sense of ordering  these execution levels can be obtained
from the  abstraction levels of language constructs  in imperative and
functional  programming  languages.   \fig{fig:prog:lang:abstractions}
summarizes them.   These levels are  similar in spirit to  the Chomsky
hierarchy  of grammars  and  their correspondence  to state  machines.
These abstraction levels also serve as execution models that gradually
transform the execution pattern towards eventual functional one.
On the  right is a  set of  levels of expression  and on the  left are
example languages that support  them.  They appear to increase upwards
from  the bottom and  reflect our  intuitive perceptions  of execution
models.  One  may either translate down  as shown by the  arrow on the
left or interpret up as shown by the arrow on the right.  The value of
these levels is  in the mechanisms offered to  express programs.  They
create name spaces (e.g.   mnemonics, variable names, function names),
state spaces (e.g.  types), automatic memory management (e.g.  garbage
collection),   and  interleaving   executions   (e.g.   closures   and
continuations). 

Given  a sense  of  ordering  execution levels,  the  three levels  of
execution for an OS have three possible orderings since the definition
of an OS fixes two of the levels.  The expression level is arbitrarily
either ``below'' the  low level model or ``between''  the low and high
level  models or  ``above'' the  high level  model.  If  an imperative
language is  used to  express an OS  that creates a  Markov algorithms
(MA) based  high level  machine using a  $\lambda$ calculus  based low
level machine,  then we  have the first  possible ordering.   This can
occur, for example, if a MA  is to be implemented over a Lisp machine,
and cross developed over a von  Neumann machine.  The OS system has to
first capture  the execution semantics  of the low level  machine, and
then use them  to express the actual bridge between  the low level and
the high level  machines as required by definition.   If an imperative
language is  used to express an  OS that creates  a $\lambda$ calculus
based  high  level machine  over  a  Turing  machine based  low  level
machine, then we have the second possible ordering.  In such cases the
model of  expression has  already bridged the  levels between  the low
level machine and itself.  The OS must bridge the expression model and
the high level one since  the high level machine defines the execution
patterns that an  OS must support.  Finally, if  a functional language
is used to express an OS  that creates a $\lambda$ calculus based high
level machine over  a Turing machine based low  level machine, then we
have the third possible ordering.  In this case the program expression
system  has already bridged  the gap  between the  low and  high level
machines, and little is needed to be bridged.  In all the three cases,
however, an OS  must express and execute the  bridge between the model
of expression  and the high level  machine, and it may  have to bridge
any gap between the model of expression and the low level machine.

Given the chosen level of program expression, an OS has to necessarily
bridge the levels execution between the language of expression and the
chosen target high  level machine.  This is expressed  in the matching
principle below and provides a mechanism to obtain the necessary steps
to reach the high level model defined for the OS.
\begin{principle}
  \label{matching:principle}
  The difference between the execution model of program expression and
  the  execution model  of the  high  level machine  is the  necessary
  component of an OS.
\end{principle}

The matching  principle is about  identifying the necessary  steps and
does  not  say  anything   about  sufficiency.   In  some  cases  more
components may be required.  For  instance, if the hardware enforces a
protected separation of kernel and  process spaces then a generic high
level language  is unlikely to support explicit  switching between the
two.   In such cases  the OS  requires the  support of  an appropriate
language  to express such  code.  Such  requirements are  specific and
vary depending  upon the definition  of the interface between  the low
level machine and the high level language.

Execution patterns  are one concern of programming  language work, and
in particular deal with closures and continuations.  Given the low and
high level  machines and the expression level,  the matching principle
helps identifying algorithms from program expression and analysis that
must  be borrowed  to realize  the high  level execution  model.  Thus
techniques of  computing closures  and continuations must  be borrowed
into the  OS problem  if the expression  model does not  support them.
The matching  principle thus  helps an OS  as an universal  machine to
figure  out what  is necessary  to bridge  between the  two  models of
computation.  However, since the  idea of ordering the abstractions in
\fig{fig:prog:lang:abstractions}  is currently  intuitive, we  rely on
insights  from programming  languages to  identify  the intermediates.
The part that is then left out for the OS problem is the algorithms to
transit from  the low  level execution model  to the higher  one.  For
example, the algorithms that remain are the ones that would eventually
create  implicit memory  for the  high level  model given  an explicit
primary memory emulating the Turing tape of the low level model.

\subsubsection{Binding and Interpretation}
\label{sec:functional:end}

An interpreter defines  the level of abstraction at  which the process
is executed.  Implemented  in software over some hardware,  it must be
seen ``raising up''  the level of interpretation.  The  source code is
the procedure.   The central construct required  for interpretation is
binding  symbols  to  their  values.   The  interpretation  system  is
discrete  and takes  one symbol  at  a time  for interpretation.   The
fundamental principle of binding is:

\begin{principle}
  \label{binding:principle}
  The binding of a value to a symbol may be established at any instant
  before the instant of interpretation.
\end{principle}

\begin{figure}[ht]
  \centering
  \epsfxsize=.5\linewidth
  \epsffile{\figdir/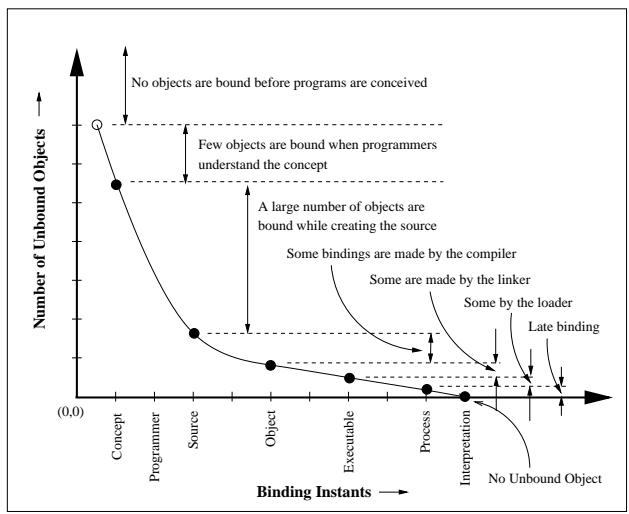}
  \caption[Temporal sequence  of bindings]{Temporal sequence  in which
    bindings are gradually established by the translation technique.}
  \label{fig:translations:binding:seq}
\end{figure}
The binding  principle is similar to  the def-use chains  in data flow
analysis that  yield compiler optimizations  like constant propagation
and common subexpression elimination.  Since interpreters are acceptor
machines,  the  binding principle  captures  the  need  to define  the
binding before  use.  Given a sequence of  symbols for interpretation,
the  principle  forms   the  base  of  the  tool   chain  approach  of
establishing binding  analogous to compilation  phases.  Bindings that
can be  computed earlier are moved  to the earlier phases  of the tool
chain.   Fig.(\ref{fig:translations:binding:seq}) captures  this idea.
The  impact  of this  principle  for  conceiving  and building  system
software  is discussed  in  another work  \cite{amv:ss}.   For the  OS
problem, this principle helps  to identify the design choices required
to map the low level machine to the high level machine.  For instance,
since framing  the main  memory is independent  of dividing  a program
into pages  it can be established  anytime before the  page table that
binds pages  to frames  is built.  Given  that the  matching principle
helps  to identify  the necessities,  the binding  principle  helps to
identify the correct possibilities of placing them.

\subsection{OSes in practice}
\label{sec:partical:oses}

Conventional OSes as a specific case of the definitions and principles
offer a ready base to verify  the proposals in this work.  First I use
the  general definition to  describe conventional  OSes as  a specific
case.  The  specific definition  and the principles  are then  used to
develop  the  core  bridging  algorithms  in two  steps:  the  general
abstract form, and their concrete instances.  The rest of the paper is
devoted to this development and check.

\subsubsection{Conventional OSes and Translation}
\label{sec:imperative:start}

Conventional OSes predominantly use the translation approach to system
development.  It  is useful to review their  overall architecture from
the point of view in this paper.  In particular, we explicitly see the
role of the binding principle.   The symbols that the CPU can directly
interpret form the machine language.  The hardware binds these symbols
to  the operations  that they  stand for.   Programmers  establish the
right set of bindings whose  interpretation does the job sought after.
However, they  rarely establish all the bindings  in machine language,
and  often a  high level  programming language  is used.   The  job is
divided into phases, and interpretation  of the result is performed in
two  different but complementary  ways.  Interpreter  techniques raise
the  level of the  raw machine  and then  directly interpret  the high
level  bindings  in this  high  level  machine.   On the  other  hand,
translation  techniques  lower  the  high level  expressions  for  the
bindings to the  machine language ones and have  the machine interpret
the lowered  set.  Fig.(\ref{fig:prog:lang:abstractions}) captures the
former technique by the upward arrow marked ``\texttt{Interpret}'' and
the latter  by the downward arrow  marked ``\texttt{Translate}''.  The
set of bindings  before the loader phase of  the translations approach
is  usually referred  to as  a  \emph{program} or  a \emph{binary}  or
simply an \emph{executable}.  A program has almost all the bindings in
machine language form.  However, a final set of bindings that describe
the  invocation time  contextual details  are left  undefined  at this
stage, and are completed at later stages like the load stage to obtain
the procedure.  This separation of the context is the basis of sharing
a  single binary  amongst  a number  of  procedures.  Instantiating  a
program essentially means completing  the bindings that have been left
unspecified in  the program and  thus building the procedure.   From a
programmer perspective, a procedure -- the conventional process -- can
be then be accurately and  constructively defined as a program with an
environment that  is sufficient for at least  one interpretation path.
Two  different invocations  of a  program,  say an  editor, differ  as
processes   only  because   of   the  different   contexts  of   their
corresponding procedures.   Defining a procedure  as a program  with a
completely  specified  interpretation  context clearly  indicates  the
activities  to  be  done  to  convert  a  program  into  a  procedure.
Subsequent interpretation of the procedure yields a process.

This view of a procedure would also hold for distributed systems where
the complete  procedure would be  composed of subsets spread  over the
components  of the  system.  For  using the  translations  approach, a
distributed system would need a  clear idea of the contextual bindings
to  arrive   at  the  notion  of  a   distributed  ``executable''.   A
distributed ``\texttt{exec()}''  would build the  required distributed
context  to   obtain  a  distributed  process   from  the  distributed
executable.  The MPI  standard uses this approach \cite{mpi:processes}
by  defining  ``an  MPI  program  is a  set  of  autonomous  processes
executing their own code, MIMD style''.

\subsubsection{Our framework for conventional OSes}
\label{sec:our:framework:for:os}

The $\lambda$ calculus  model is a good high level  machine for an OS.
It is an  intuitively simple view for an end user  since it strips off
all  the  unnecessary details.   After  all,  an  algorithm needs  the
ability  to  identify  its  distinct  components,  prescriptions  that
describe transformation of some objects into others, and an ability to
apply a prescription to the components of the algorithm.  The rules of
the  $\lambda$ calculus capture  this intuition.   A computer  user is
typically  concerned  with  obtaining  the results  of  applying  some
transformation  on  some  objects.   Technological  details  like  the
representation of the components,  their storage and retrieval, or the
variety of ways in which they may be made to interact via applications
are  not relevant to  the end  user.  The  $\lambda$ calculus  view of
computation can be  the candidate high level machine  if the technical
details  are   hidden  away   through  techniques  like   file  format
standardization  and implicit  memory  management.

A practical  OS is built over  low level Turing  machine like hardware
with  an  empirically  defined  higher  level.   Using  the  $\lambda$
calculus as this high level machine, a conventional OS in practice can
be  better described  as a  software  that wraps  an imperative,  i.e.
Turing machines based, technology  under a functional, i.e.  $\lambda$
calculus, based front.  In other words, it creates an abstract machine
on which end users ideally  can execute programs in a functional style
over an  imperative hardware.  OSes in practice  are almost functional
style  with  sophisticated  naming   via  icons  and  application  via
drag-and-drop  capabilities.  The  compositional  capabilities of  the
Unix system  are a  persuasive example of  this view.  The  user level
view    and    the    kernel    level    views    in    Yates    \etal
\cite{Yates99i/oautomaton} respectively  could correspond to  the high
level  almost functional  machine  (the  user view)  and  a low  level
imperative  machine  (the  kernel  view).   With  this  view  suitable
resource management is a  step towards achieving the abstract machine,
and the  abstract machine  to aim for  is precisely a  functional one,
i.e.  as described by the $\lambda$ calculus.  In practice, OSes still
have  some way  to go  to reach  full capabilities  of  the functional
approach.  For instance, anonymous  $\lambda$ expressions are still to
be  realized, and  still more  simply not  all programs  return values
called ``exit codes'' to the OS.

\subsubsection{The problem of a conventional OS}
\label{sec:usual:os:prob}
\label{sec:os:abstractions}

A  conventional  OS  can be  defined  as  an  algorithm that  aims  to
implement the $\lambda$ calculus model  of computation in terms of the
Turing machine model.  To realize  the $\lambda$ calculus model over a
Turing machines based hardware, an  OS needs to identify the essential
abstractions.  For the high  level machine $\lambda$ calculus requires
three  abilities: naming, function  abstraction, and  application.  At
the low  level we are  given an infinite  countable tape, a  head that
captures the state transition abilities,  and the basic read and write
operations on the tape.  The imperative expression style for low level
machine must  be used to  express and eventually realize  a functional
execution style.  The program of this section is to use the principles
and definitions  to identify  the essential abstractions  required for
this  purpose.   Some   abstractions  are  borrowed  from  programming
languages,  and  some  must   be  constructed.   The  details  of  the
abstractions borrowed  from other areas will be  considered well known
enough to allow focus on the constructed ones in the rest of the work.

Analogous to  the abstraction levels in programming  languages, we can
conceive  abstraction  levels  for  such  an  OS  that  implement  the
execution model  of a  high level functional  machine in terms  of the
execution model  of a low  level imperative one.  Given  an imperative
hardware,  we  posit  that  the   OS  uses  the  usual  algorithms  to
(eventually) realize  a functional machine,  and hence one  can expect
the abstraction levels to map to the usual algorithms.
\begin{figure}[t]
  \centering
  \epsfxsize=.6\linewidth
  \epsffile{\figdir/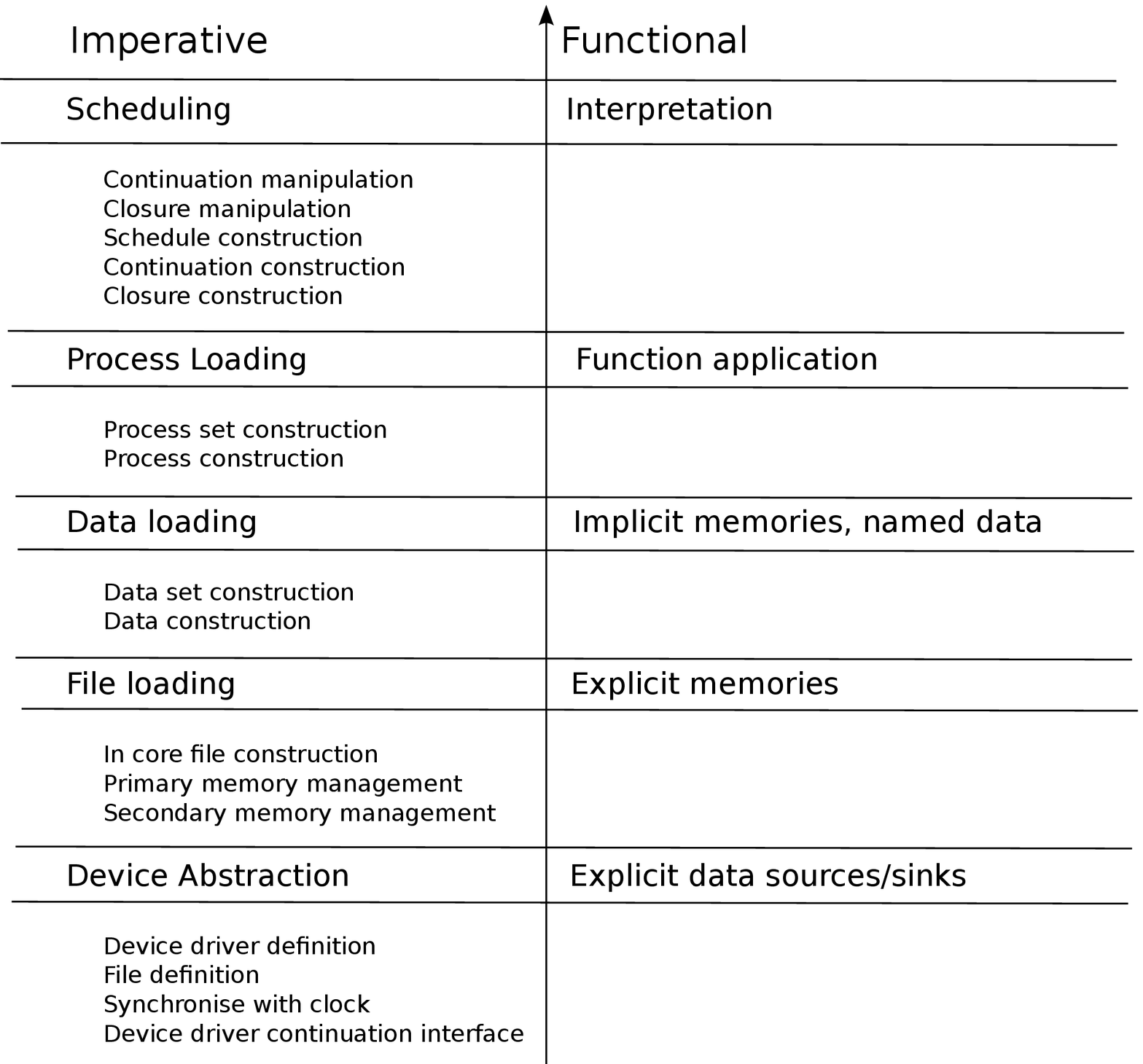}
  \caption[Abstractions in an OS]{Abstractions in an OS that implement
    a high level functional machine over a low level imperative one.}
  \label{fig:os:abstractions}
\end{figure}
A  functional  machine   needs  three  essential  mechanisms:  naming,
function abstraction, and application.  Abstraction is usually the end
result  of program  expression  and  may be  ignored  for the  present
purpose.     

Fig.(\ref{fig:os:abstractions}) shows  the gradual realization  of the
functional machine  for execution.  The  stages of realization  are on
the left, and  the intermediates of the functional  machine are on the
right.  Reading it  upwards from the bottom as  suggested by the arrow
in  the middle, an  OS first  deals with  I/O devices  particularly to
synchronize  the input with  the system  clock and  bring it  into the
system as data.   A continuation, usually an interrupt,  is used.  The
primary memory is the final target of the data.

A  name  binds  a  symbol  to  an object  in  the  system.   Within  a
conventional OS procedures are  named using natural numbers, and files
are  named using  strings.  Strings,  in turn,  are  represented using
natural  numbers.   Other ways  of  naming  files  are through  memory
addresses  (e.g.  register  names, memory  location  values, functions
that  compute  memory  location  values called  pointers),  I/O  ports
(hardware  or  software enumerated),  device  names  (since these  are
sources or sinks of data streams), and even through compositions where
the file is a set of sets of natural numbers -- a data structure.  The
choice  of  the  data  type  used  to name  is  an  issue  of  program
expression, and the  essential principle is naming the  entity.  An OS
uses a number  of sets of various types as a  storehouse of names that
it can use to bind objects.

Data  operations can be  accomplished via  I/O devices  like secondary
memories and networks.   Any secondary memories that may  exist in the
system  are also  made amenable  to load/store  operations  using file
system techniques.   In particular, file systems bind  (path) names to
data objects, and load/store operations  use the names to transfer the
data objects from/to primary memory.   In the primary memory, the data
objects  are   usually  named   using  natural  numbers   called  file
descriptors.  The usual files thus exist in the primary memory as sets
of natural  numbers named using file descriptors.   This completes the
in-core  representation of files  as a  set of  named sets  of natural
numbers.  Data operations can also be accomplished by binding the data
objects to code  that generates the value.  For  example, IP addresses
are obtained by querying name servers during execution.

Data loading is  complete when a set of named  sets of natural numbers
required for later  interpretation is ready.  The data  object is said
to have  been instantiated.  Until  this point, memory  management for
object instantiation is often  an explicit part of program expression.
Explicit memory management coupled with loading operations is required
for instantiation.   However, the  binding principle asserts  that the
only requirement is that the  data load operations finish before their
use in interpretation.  The  object instantiation can hence be delayed
to  later than program  expression.  In  other words,  implicit memory
techniques are now  possible, and programs can be  expressed in a more
functional style.

A function  application is simply expressed as  program with arguments
for  operation.  In  response, an  OS loads  the program  into primary
memory  as a  file, completes  any contextual  bindings  required, and
obtains a set  of natural numbers that can be  scheduled.  This is the
conventional process.  In  general, an OS deals with a  set of sets of
natural numbers.  Some sets in  this set are from data load operations
and others are from process load operations.  At this point a function
application is ready to run.

To run a function application, a  CPU time schedule must be fixed.  An
OS  has a  set of  one or  more applications  to run,  and a  CPU time
allocation  schedule  for  each.    To  switch  between  the  function
applications an OS computes the  closure to capture the data state and
the  continuation  to capture  the  interpretation  state.  These  may
additionally be manipulated too.  For example, an \texttt{exec()} call
overwrites the closure and continuation of the parent with its own.

It  is  useful  to observe  that  although  the  program to  create  a
functional  style high  level  machine over  an  imperative low  level
machine appears to be in place,  it is not so.  First classness is far
from complete at all levels, and most current OSes distinguish between
``executables''  and ``(data) files'',  particularly for  security and
trust reasons.   It is difficult  to express and execute  higher order
procedures,  anonymous  procedures   etc.   Lazy  evaluation  is  also
partial, e.g. Unix pipes.

\section{Conventional OS algorithms -- symbolic forms}
\label{sec:algorithms}
\label{sec:conventional:os:algorithms}

To place the core algorithms of conventional OSes I start bottom up by
searching for basic operations over  the low level machines that would
be needed to realize the OS.  The motivations for the basic operations
over such machines also suggest the set of resources to use.  I evolve
a  symbolic  machinery  to  express  the algorithms  using  the  basic
abstract  operations.  The implicit  assumptions in  these expressions
are  used  to  examine  the  possible  concrete  realizations  of  the
algorithms.  While the  attempt is to be as  orthogonal as possible, I
make  no effort  to establish  the formal  properties of  the symbolic
system.  That would be a part of the formalization effort.

As  an universal  machine, an  OS starts  with a  set of  one  or more
procedures that must be executed.  Disjoint subsets of files that form
procedures  and  the  resources  in  the low  level  machine  must  be
constructed  so that  they can  be bound  to each  procedure.   From a
Turing  model von  Neumann  architecture perspective,  there are  four
issues that  an OS has to deal  with: (a) deciding to  which subset of
cells of the tape should the transition function be applied to and for
how long,  (b) organize  the contents  of the cells  of the  tape, (c)
communicate to and fro with the  outside world and the tape cells, and
(d) ensure  that the  subsets remain disjoint.   Conventionally, these
are respectively  known as process scheduling,  memory management, I/O
management, and  protection.  There are essentially  only two physical
resources, the CPU time and the  memory space that need to be managed,
and these  two resources  are the parameters  of a set  of algorithms.
Two more sets  of logical, or conceptual, resources  are required: the
set of names -- the first  of the three mechanisms required for a high
level  functional machine,  and the  set  of states.   Subsets of  all
resources  must be  bound to  each procedure  of an  OS.   The logical
resources are the glue that bring the system together.  To ensure that
the resources are maximally utilized and divided into disjoint subsets
we induce an equivalence class partitioning of each resource set.

For a single procedure, i.e.  a single set of natural numbers, on such
a computer  system the algorithm  is simple: Allocate all  the primary
memory and  the CPU time  to the procedure  after loading it  from the
secondary device.  The procedure and file naming need not be explicit.
The next step is to devise algorithms for a set of more than one files
each of which would be a  procedure at some later time instant.

\subsection{Basic definitions}
\label{sec:basic:defns}

A nonempty set  of natural numbers, $p$, is the  basic unit of concern
for an OS.   $p$ is a file  until it consumes CPU time.   Let $\#_s: p
\rightarrow \mathbb{N}$ be the  number of memory units associated with
a file $p$, and called  as the spatial cardinality of $p$.  Similarly,
let $\#_t: p  \rightarrow \mathbb{N}$ be the number  of CPU time units
associated with a file $p$,  and called as the temporal cardinality of
$p$.   $\#_t(p)$,  the length  of  the  computation  generated by  the
natural  numbers in  $p$, is  in  general undecidable.   Also, $\#:  S
\rightarrow \mathbb{N}$  is the  usual cardinality of  any set  $S$ as
usual.  The file $p$ has to  be accommodated in memory and may have to
be scheduled on the CPU.  A set $P$
\begin{equation}
  \label{eq:basic:set:of:files}
  P \names \{p_1, p_2, p_3, \ldots\}
\end{equation}
of such files is manipulated by an OS, i.e. resources are allocated to
each member in $P$ (the notation ``$\names$'' introduces the symbol on
its LHS as the name of the object on its RHS).

\subsection{Universal OS operations}
\label{sec:universal:os:functions}

The OS algorithms  are the means to realize the  high level machine of
an OS.   The question is:  are there a  set of core  useful operations
that  form the  backbone of  OS algorithms?   This section  offers the
arguments for their existence and identifies the operations.

A UTM  has tape cells  and discrete time  instants as resources  to be
utilized  and  provides the  basic  \texttt{read}, \texttt{write}  and
\texttt{direction} operations.  An individual cell or time instant has
to be \emph{selected} by an OS for use by one of its procedures.  This
requires that each cell  or time instant be uniquely \emph{enumerated}
for addressing when required.  A  \emph{set} of cells or instants must
be  collected  together  to   be  associated  with  a  procedure.   To
manipulate a  superset (like $P$  in Eq.(\ref{eq:basic:set:of:files}))
of  such  sets  OS  algorithms  need operations  to  \emph{append}  or
\emph{remove}  a given  set  to  or from  the  superset.  And  finally
\emph{bind} or  \emph{unbind} operations are  required to respectively
associate or  de-associate a set to  the tape or the  head, or another
set.   The working  hypothesis of  the current  work is  that specific
computable instances of these  operations are used to realize specific
algorithms  for  an  OS.   Section \ref{sec:manage:algos}  uses  these
operations  to  formally  express   the  OS  algorithms,  and  section
\ref{sec:algorithms:concrete} identifies  the implicit assumptions and
relaxes them to explore the design space for specific algorithms.  The
set of operations required to  describe OS algorithms are (assume that
all the  operations return $false$ on failure,  capital letters denote
arbitrary sets, and $\mathbb{N}$ is the set of natural numbers):

\begin{enumerate}
\item  For any set  $S$ and  given another  set $\mathcal{N}$  of some
  elements such that $\#\mathcal{N} \ge \#S$, an enumeration function,
  $Enum:  S,  \mathcal{N} \rightarrow  S$,  associates  an element  of
  $\mathcal{N}$,  called the  \emph{address}, to  each element  of set
  $S$.  A  successful $Enum$ enables addressing  individual members of
  $S$, and $S$ is  said to be \emph{addressable} or \emph{enumerated},
  and  is a finite  function for  finite reusable  resources.  Usually
  $\mathcal{N}  = \mathbb{N}$, but  can be  any enumerable  data type.
  For example,  the register file  within a CPU  is a finite  set, and
  $\mathcal{N}$  is  a finite  set  of  strings called  \emph{register
    names}.   Assume $\mathcal{N}  = \mathbb{N}$  in the  rest  of the
  work.
\item  A selection  function,  $Sel :  P,  \mathbb{N} \rightarrow  p$,
  extracts some  $i^{th}, i  \in \mathbb{N}$ member  $p$ from  the set
  $P$.
\item An organizing function, $Org  : P \rightarrow P$, rearranges the
  members in $P$.   
\item The $bind: o_1,o_2 \rightarrow b$ operation takes two objects of
  arbitrary types  $o_1$ and  $o_2$ and binds  them into an  object of
  type pair $b$.
\item The  $unbindfst: b \rightarrow  o_1$ operation takes a  pair $b$
  and returns the first object of the pair.
\item The  $unbindsnd: b \rightarrow  o_2$ operation takes a  pair $b$
  and returns the second object of the pair.
\item The $append: L,o \rightarrow L$  takes an object of type set $L$
  and an object of  arbitrary type $o$ and adds $o$ to  $L$ to yield a
  set with one more element.  $append$ can be realized in terms of the
  $bind$ operation.
\item Inversely, $remove:  L,o \rightarrow L$ takes an  object of type
  set $L$ and an object of arbitrary type $o$ and removes $o$ from $L$
  and returns an object of type set $L$ with the object $o$ removed.
\item  The  $makeset: S,  \mathbb{N}  \rightarrow  S$  that returns  a
  contiguous subset  of resource  set $S$ with  size specified  by the
  second argument.
\end{enumerate}
Table \ref{tab:os:functions} summarizes these.
\begin{table}[t]
  \begin{center}
    \begin{tabular}{|l|l|l|}
      \hline 
      \textbf{Name} & \textbf{Signature} & \textbf{Description} \\
      \hline 
      Select & $Sel : P, \mathbb{N}  \rightarrow p$ &
      extracts $i^{th}$ member $p$ of $P$ \\
      Organize & $Org : P \rightarrow  P$ & rearrange the members in $P$ \\
      Enumerate &  $Enum:  S, \mathcal{N}   \rightarrow  S$ &
      associate  an element  of  $\mathcal{N}$ to  each element of set $S$ \\
      Makeset & $makeset: S, \mathbb{N} \rightarrow S$ & 
      \parbox[t]{.4\textwidth}{return a
        contiguous subset of $S$ with size specified by the second
        argument} \\
      Bind & $bind: o_1,o_2  \rightarrow  b$ & binds objects $o_1, o_2$
      into a pair $b$ \\
      Unbindfst & $unbindfst: b \rightarrow  o_1 $ & unbinds a pair $b$
      and returns first object $o_1$ \\
      Unbindsnd & $unbindsnd: b \rightarrow  o_2 $ & unbinds a pair $b$
      and returns second object $o_2$ \\
      Append & $append: L,o  \rightarrow L$ & appends an object $o$ to a
      set (or list) $L$ \\
      Remove & $remove: L,o  \rightarrow L$ & removes an object $o$ from
      a set (or list) $L$ \\
      \hline 
    \end{tabular}
  \end{center}
  \caption[Functions necessary for an OS]{The functions necessary for
    symbolic expression of OS algorithms (see section
    \ref{sec:universal:os:functions}).}
  \label{tab:os:functions}
\end{table}

\subsection{Notation}
\label{sec:notation}

In the  following, the  linear notation for  predicates is  taken from
Gries and Schneider  \cite{Gries:1993:LAD:161182}.  The usual notation
like:  $\sum_{i =  1}^{i  < N}  f(i)$  looks like:  $\left  (+\ i  \in
  \mathbb{N} \vert (1  \le i < N) : f(i)\right)$.  On  the left of the
vertical  bar we  specify  operation being  quantified,  $+$, and  the
general predicates for properties  of the quantified variables, $i$ in
this case.  On  its right before the ``:''  we specify the predicates,
$(1 \le  i < N)$, that  the quantified variables must  satisfy for the
given expression.   After the ``:'' we specify  the actual expression,
$f(i)$, over  which the quantification is  being defined.  Parenthesis
serve  to group expressions  as usual.   The expressions  are visually
structured  as  Lisp S-expressions.   I  also  use the  ``\texttt{Let}
$\langle$bind   exprs$\rangle$   \texttt{in}  $\langle$body$\rangle$''
construct  similar  to the  \texttt{let}  syntax  sugar  in Lisp.   It
explicitly indicates that the  ``\texttt{bind exprs}'' objects must be
bound before  the expressions in the  ``\texttt{body}'' are evaluated.
The construct  captures the use of  the binding principle  to build an
algorithm.

\subsection{Resources and Subsets}
\label{sec:resources:subsets}

There  are two  kinds of  resources: physical  and  logical.  Physical
resources are hardware components like clock and memory, while logical
components  are  conceptual artifacts  like  names  and states.   Some
resources may  be finite  in practice, e.g.  memory.  They  are finite
since the processes that  produce them have terminated.  Resources are
also characterized as  being \emph{reusable} if they can  be used over
and over,  or being \emph{consumable} if  they can be  used only once.
Names  are  logical  reusable  resources  and CPU  time  instants  are
physical consumable resources.

Resources are described  as the sets: memory $M$,  CPU time $T$, Names
$N$ and System states $\Sigma$:
\begin{eqnarray}
  \label{eq:the:memory:set}
  M &\names& \{m_0, m_1, m_2, \ldots m_{(L-1)}\} \\
  \label{eq:the:name:set}
  N &\names& \{n_0, n_1, \ldots\} \\
  \label{eq:the:cputime:set}
  T &\names& \{t_0, t_1, \ldots\} \\
  \label{eq:the:state:set}
  \Sigma &\names& \{\sigma_0, \sigma_1, \ldots\}
\end{eqnarray}

\eqn{eq:the:memory:set}  is  finite in  practice  and  is  a set  with
cardinality   $\#M   =   L$   of   a   physical   reusable   resource.
\eqn{eq:the:name:set} is a countably  infinite set of logical reusable
resources.   In  contrast   \eqn{eq:the:cputime:set}  is  a  countably
infinite physical  consumable resource.  However, $\Sigma$  would be a
part  of the OS  structure that  deals with  the execution  models, in
particular the  computation of closures and continuations.   It is not
considered in this  work since the goal is to  manipulate $M$, $N$ and
$T$ to implement  the definitions of an OS.  Subsets  of each of these
sets are, or are to be, associated  with every $p \in P$.  A subset of
$M$ is called  a \emph{(memory) region} and a subset  of $T$ is called
an \emph{interval}.

To ensure  disjointness, subsets in  OS for our simple  computer model
have the property that an element  belongs to one and only one subset,
i.e.  for any set $S$ of resources $s_i$ and a set $U \subseteq S$, we
have
\begin{equation}
  \label{eq:subset:prop:01}
  \biggl(\forall i \in \mathbb{N}\ \vert\ (s_i \in S)\ :\ \bigl(s_i
  \in U\bigr) \equiv \bigl(s_i \not \in (S - U)\bigr) \biggr)
\end{equation}
Some  resources may  be enumerated  using natural  numbers.  A  set $U
\subseteq  Enum(S,  \mathbb{N})$  of  such  addressable  resources  is
described by  two natural  numbers, $u_{min}$ and  $u_{max}$, $u_{min}
\le u_{max}$, and is constructed as:
\begin{eqnarray}
  \label{eq:subset:desc:01}
  U &\names& (\exists u_{min}, u_{max} \in \mathbb{N}\
  \vert\ (u_{min} \le  u_{max}) \wedge (s_j \in Org(Enum(S)))\ :
  \nonumber 
  \\ &~& \quad\quad
  (\forall j \in \mathbb{N}\ \vert (u_{min} \le j \le u_{max})\ :
  \nonumber \\ 
  &~& \quad\quad\quad
  Sel(Org(Enum(S,\mathbb{N})), j) ))
\end{eqnarray}
At  the hardware  level $Enum(M,\mathbb{N})$  and $Enum(T,\mathbb{N})$
are  simple  enumeration  functions  and  $Org(M)$  and  $Org(T)$  are
identity  functions implemented  by  the memory  addressing and  clock
mechanisms respectively for a  single computer system.  These are more
complex for distributed systems since while $Enum$ may be defined over
the  entire system,  $Org$ may  typically  be defined  to exploit  the
communication    characteristics.    \eqn{eq:subset:desc:01}   creates
subsets of  addressable resource $S$ as  required.  The \emph{makeset}
function in  \eqn{eq:subset:function} converts this  into an algorithm
that accepts the set of  resources, e.g.  $M$, and desired cardinality
$r$ as arguments.
\begin{eqnarray}
  \label{eq:subset:function}
  makeset (S,\ r) &\names& 
  \biggl(\exists u_{min}, u_{max} \in \mathbb{N}\ \vert\ (u_{min} \le
  u_{max}) \wedge ((u_{max} - u_{min}) = r)\ :
  \nonumber \\ &~&
  \quad\quad\quad
  \biggl(\forall j \in \mathbb{N}\ \vert\  (u_{min} \le j \le u_{max}) 
  \wedge (s_j \in Org(Enum(S,\mathbb{N})))\ :  
  \nonumber \\ &~&
  \quad\quad\quad \quad\quad\quad
  Sel(Org(Enum(S,\mathbb{N})), j) 
  \biggr)\biggr)
\end{eqnarray}
Partial evaluations of  \eqn{eq:subset:function} for specific resource
--  $M$ or $T$  -- yield  subset constructors  over that  resource and
accepts the size required as an argument.  $(makeset\ \ T)$ constructs
subsets  of the  CPU time  resource  and $(makeset\  \ M)$  constructs
subsets of the memory resource.  The partial evaluations show that the
algorithmic  structure  is  independent  of  the  physical  resources.
Reusability  of  a  resource  implies  that  an  ``\emph{releaseset}''
operation is possible while the  finiteness of a resource implies that
it is required.  No such operation can exist for consumable resources.

Let $R$ denote the subset created by \eqn{eq:subset:function}, i.e.
\begin{eqnarray}
  \label{eq:mem:subset}
  R &\names& makeset\ (S,\ r)
\end{eqnarray}
IF $S \equiv M$, then $R$ is a \emph{memory region}.  It $S \equiv T$,
then $R$  is a \emph{time interval}.   Let $p_k \names  Sel(P, k), p_k
\in P$  be such that $r_k  \names \#p_k$ so that  $R_k \names makeset\
(S,\ r_k)$.  Let $b_k$ be the pair obtained by binding $p_k$ to $R_k$,
\begin{eqnarray}
  \label{eq:mem:bind}
  b_k 
  &\names& bind(p_k, R_k)
  \nonumber \\
  &\equiv& bind(Sel(P, k),\ makeset\ (M,\ \#_s(Sel(P, k))))
\end{eqnarray}
For any  set $S$,  with subsets $S_O$,  the ``occupied''  subsets, and
$S_F$, the ``free'' subsets, such that  $S_O \cap S_F = \phi$ and $S_O
\cup S_F = S$, we will have the following invariants:
\begin{eqnarray}
  \label{eq:total:size:invariance}
  \#S 
  &=& \#S_O + \#S_F \nonumber \\
  &\equiv& 
  \underbrace{\#\bigl(\cup \ i \in \mathbb{N}\ \vert\ (R_i \in S_O)\
    :\ R_i \bigr)}_{\#S_O}  +
  \underbrace{\#\bigl(\cup\ i \in \mathbb{N}\ \vert\ (R_i \in S_F)\ :\
    R_i \bigr)}_{\#S_F} \\ 
  \label{eq:single:membership}
  \bigl(\forall i \in \mathbb{N}\ &\vert& (R_i \in S_O) \not \equiv (R_i
  \in S_F)\bigr) \\
  \label{eq:union:is:total:set}
  M &=& \bigl(\cup i \in \mathbb{N}\ \ \ \vert\ \ \ ((R_i \in S_O) \vee
  (R_i \in S_F))\ : R_i\bigr) \\
  \label{eq:disjointness}
  \bigl(\forall i,j \in \mathbb{N}\ &\vert& (i \not = j)\ :\ (R_i \cap
  R_j) = \phi\bigr)
\end{eqnarray}
\eqn{eq:total:size:invariance} says that the  total size of $S$ is the
sum    of    the   sizes    of    the    sets    $S_O$   and    $S_F$.
\eqn{eq:single:membership} says  that an element  of $S$ is  either in
$S_O$ or  in $S_F$  but never both.   \eqn{eq:union:is:total:set} says
that  the union  of all  the subsets  is the  total set.   And finally
\eqn{eq:disjointness}   expresses  that   all  subsets   are  mutually
disjoint, i.e. a given element of $S$  is a member of one and only one
subset.    The   subset  operation   induces   an  equivalence   class
partitioning  of  $S$.   This  isolation  of the  resource  subset  is
necessary to protect a file from others.

For  a resource set  $S$ enumerated  using set  $N$, the  algorithm to
obtain the subsets of $S$ for each element of $P$ is:
\begin{eqnarray}
  \label{eq:basic:mgmt}
  &~&
  \biggl(
  \forall i \in \mathbb{N}\ \vert\ (1 \le i \le \#P)\ :\ 
  \nonumber \\
  &~& \quad \quad 
  \bigl(\mathrm{Let}\ (p_i \names Sel(Org(Enum(P,N)), i)) \wedge (r_i
  \names \#p_i) \wedge (\#S_F \ge r_i)
  \ \mathrm{in} 
  \nonumber \\
  &~& \quad \quad \quad \quad (\mathrm{Let}\ (R_i \names makeset(S_F,
  r_i)) \ \mathrm{in} 
  \nonumber \\
  &~& \quad \quad \quad \quad \quad \quad remove (S_F, R_i) \wedge
  append(S_O, R_i)  \wedge bind (p_i, R_i) 
  )\bigr)\biggr)
\end{eqnarray}
and starts  with the state:  $(S_O = \phi)\  \wedge\ (S_F =  S)$.  The
invariant  is: $(S_O  \cap S_F  = \phi)  \wedge (S_O  \cup S_F  = S)$.

\eqn{eq:basic:mgmt} describes a  set of bindings of subsets  of $S$ to
files in set $P$ and is  type consistent.  It says: for each member of
$P$, extract the member from  $P$ ($Sel(P, i)$), get its size $\#p_i$,
check if $S$ has enough  elements available ($\#S_F \ge \#p_i$), carve
out a  suitable subset $R_i$ ($makeset(S_F, r_i)$),  remove $R_i$ from
the  set $S_F$,  add it  to the  set $S_O$,  and bind  $R_i$  to $p_i$
($bind(p_i, R_i)$).   The principle of  binding occurs in two  ways in
\eqn{eq:basic:mgmt}.  In  an expression like  $Sel(Org(Enum(P,N)), i)$
it only requires that the  arguments of an outer operator be generated
before its  own evaluation.  The  $Let$ expression also  requires that
its ``\texttt{bind exprs}''  be evaluated before the ``\texttt{body}''
expression.  For  example, $Enum$  is predefined and  pre-evaluated by
the addressing scheme in hardware if  $S$ is the memory, or is defined
by the  system clock  if $S$ is  the CPU  time resource.  Any  term in
\eqn{eq:basic:mgmt} that is  a set, e.g.  $p_i$, can  be operated upon
by suitable $Enum$  and $Org$ functions.  We have  shown it explicitly
for the set $P$.
 
The set of bindings processed by predicate \eqn{eq:basic:mgmt} is:
\begin{eqnarray}
  \label{eq:basic:mgmt:set}
  L &\names& \biggl\{
  \forall i \in \mathbb{N}\ \vert\ (1 \le i \le \#P)\ \wedge\  
  \nonumber \\
  &~& \quad \quad 
  \bigl(\mathrm{Let}\ (p_i \names Sel(Org(Enum(P)), i)) \wedge (r_i
  \names \#_sp_i) \wedge (\#S_F \ge r_i)
  \ \mathrm{in} 
  \nonumber \\
  &~& \quad \quad \quad \quad (\mathrm{Let}\ (R_i \names makeset(S_F,
  r_i)) \ \mathrm{in} 
  \nonumber \\
  &~& \quad \quad \quad \quad \quad \quad remove (S_F, R_i) \wedge
  append(O, R_i) \wedge b_i \names bind (p_i, R_i)
  )\bigr)\ :\ b_i\biggr\}
\end{eqnarray}

\subsection{The resource manipulation algorithms}
\label{sec:manage:algos}

The algorithms  in section \ref{sec:resources:subsets}  can be applied
to the two physical resources of an  OS, using the names in $N$ as the
glue.  We choose suitable sets, $N$, to name the various components of
the   algorithm  to   realize   the  high   level  machine.    Section
\ref{sec:basic:name:management} looks  at the basic  algorithm to bind
names to objects.  The algorithms for memory resource and the CPU time
resource  can  be obtained  by  applying  the  expressions in  section
\ref{sec:resources:subsets} to the respective sets.

\subsubsection{Names -- the glue}
\label{sec:basic:name:management}

Given a countable  set $N$, define a function  $name: P, N \rightarrow
B$ that takes  a elements $n \in N$,  $p \in P$ and returns  a pair $b
\names (n, p),\ b \in B$ that binds  a name $n$ to a file $p$ iff $\#N
\ge  \#P$.  In  general, $name$  can be  many-to-one function,  but is
often one-to-one.  $N$ is $\mathbb{N}$ when $P$ is a set of processes,
the members of $N$  are called \emph{process identifiers}, and $names$
is a one-to-one function.   $N$ is a set of strings when  $P$ is a set
of files, the members of $N$ are called \emph{file names}, and $names$
can be  a many-to-one function since  the uniqueness of  names is only
required between the CPU and the primary memory.  The naming operation
is the predicate:
\begin{eqnarray}
  \label{eq:basic:naming}
  &~&
  \biggl(
  \forall i \in \mathbb{N}\ \vert\ (1 \le i \le \#P)\ :\ 
  \bigl(\exists j \in \mathbb{N}\ \vert\ (1 \le
  j \le \#N)\ : 
  \nonumber \\
  &~& \quad \quad \quad \quad \quad \quad bind (Sel(P, i), Sel(N, j)) 
  \wedge N \names remove (N, Sel(N, j)) 
  \bigr)\biggr),
\end{eqnarray}
and the set of bindings is given by \eqn{eq:basic:mgmt:set} for memory
resource.
\eqn{eq:basic:naming} ensures  a one-to-one mapping  by removing every
bound name from  the set of names $N$.   Ignoring the remove operation
can yield a possibly many-to-one mapping.

The naming operation is used to  associate a member of $P$ to a subset
of $M$  for load/store operations, to  a subset of  $T$ for scheduling
operations, to a subset of $\Sigma$ for closure operations, and all of
these bindings together  form a state of the  process corresponding to
the     member     of      $P$.      The     \emph{name}     operation
(Eq.(\ref{eq:basic:naming})) is the mechanism to establish abstraction
levels since it is the glue that binds physical resources to processes
of individual procedures.  It implements the ``$\names$'' notation for
concrete  algorithms,  and  will   be  understood  implicitly  in  the
following discussion.

\subsubsection{Memory allocation and deallocation}
\label{sec:basic:mm}

Memory  allocation, or  binding a  region to  a file,  is  required to
develop implicit memory techniques to reach the high level destination
machine of an OS.  Finiteness of $M$ requires a deallocation operation
defined on $M$.

The memory  management problem is identical for  primary and secondary
memories.  For memory, we have $S  \equiv M$.  The subset $S_O$ is the
set of  memory occupied by some files,  and $S_F$ is the  set of free,
available memory.   The set $L$ is a  set of all bound  pairs $b_i$ of
files $p_i$ loaded at region $R_i$.  It corresponds to data structures
like page and segment tables.

Since memory  is a finite resource,  the predicate $(\#F  \ge r_i)$ in
\eqn{eq:basic:mgmt:set} that  checks for availability  of the required
amount of  resource may not  be always true,  and has to  be evaluated
every  time.  However, being  a reusable  resource, the  memory system
admits the $unbind\star$ operations.  We use the $unbindsnd$ operation
to dissociate  a pair $b_i \names  (p_i,\ R_i)$ and  return the region
$R_i$.  The \emph{releaseset} predicate is given by
\begin{eqnarray}
  \label{eq:basic:mm:deallocation}
  releaseset(b_i) &=& 
  \biggl(\mathrm{Let}\ (R_i \names unbindsnd(b_i))\  \ \mathrm{in}\ \ 
  \nonumber \\
  &~&\ \ \ \ \ \ \bigl(remove (O, R_i) \wedge append(F, R_i) \wedge
  remove (b_i, L)  \bigr)\biggr)
\end{eqnarray}
\eqn{eq:basic:mm:deallocation}  uses  $unbindsnd$  and enables  memory
reuse.  The higher level machine we defined would need memory to be an
implicit  component.    \eqn{eq:basic:mm:deallocation}  is  the  basic
requirement  of such  a goal.   The principle  of binding  can  now be
employed  to decide on  a time  instant at  which an  automatic memory
allocation-deallocation  algorithm  would   employ  memory  reuse  via
\eqn{eq:basic:mm:deallocation}.

\subsubsection{CPU time allocation}
\label{sec:basic:cpu:time}

The basic operation  for managing CPU time is again  the creation of a
subset of the required size.  We have $S \equiv T$, $S_O$ would be the
CPU  time intervals that  have been  utilized, and  $S_F$ would  be the
available  CPU  time  intervals.   Since  CPU  time  is  a  consumable
resource,  the predicate  $(\#S_F \ge  r_i)$  in \eqn{eq:basic:mgmt:set}
that checks  for availability  of the required  amount of  resource is
always true.

$L$ in \eqn{eq:basic:mgmt:set} for the CPU  time is a set of all bound
pairs $b_i$ of files $p_i$ scheduled for execution.  It corresponds to
data structures of the scheduler.

\section{A concrete algorithm for a conventional OS}
\label{sec:algorithms:concrete}

Section \ref{sec:algorithms} looked at the basic algorithms that an OS
uses symbolically.  It identified  some primitive functions that an OS
uses  and  used  them  to  obtain  constructive  expressions  for  the
algorithms.  This  section develops a specific  algorithm by examining
the assumptions  implicit in the simple  constructions above, relaxing
them, instantiating  suitable $Enum$,  $Org$, and $Sel$  functions and
combining  the instances  in various  ways.  These  functions together
with    the    predicates    and    set   expressions    of    section
\ref{sec:algorithms} can offer a framework to develop new algorithms.

\subsection{Implicit assumptions in symbolic algorithms}
\label{sec:algorithms:concrete:assumptions}

A number of assumptions are implicit behind the expressions in section
\ref{sec:algorithms}.  The set $P$ (Eq.(\ref{eq:basic:set:of:files})),
its members  $p_i$, and the resources  $M$ and $T$ can  be examined to
systematically extract the implicit assumptions.

We have  assumed that the set $P$  is given, and does  not change with
time once  given.  An  OS where  the $P$ is  given beforehand  even if
arbitrarily, is a \emph{batch processing OS}.  If the set $P$ is to be
updated dynamically as and when needed, the OS acquires the capability
to be  \emph{interactive}.  An incoming procedure can  either be added
at the beginning of the set, or at the end of the set, or somewhere in
between.  An  OS may treat the set  $P$ as a \emph{queue},  and add an
incoming  procedure  to the  instantaneous  end  of  the queue  unless
specified   otherwise.    Conventional  OSes   implement   $P$  as   a
\emph{circular queue}.   

For  each member,  $p_i \in  P$,  we have  assumed that  the CPU  time
($\#_tp_i$)  and  memory  space  requirements  ($\#_sp_i$)  are  given
beforehand, and  do not change.  However $\#_tp_i$  is undecidable and
$\#_sp_i$   can  change   with  time.    Let  a   computable  function
$f_i(\#_tp_i)$ denote the change in memory requirements of a procedure
$p_i$.  Let
\begin{equation}
  \label{eq:dynamic:mem}
  \delta m_i = (f_i(k + 1) - f_i(k)),\ \ \ \ \ \mathrm{for\
    some}\ k \in \mathbb{N},\  1 \le k \le (\#_tp_i - 1).
\end{equation}
If $\delta m_i > 0$, then the file increases and we use $makeset$.  If
$\delta m_i < 0$, then the file decreases and we use $releaseset$.  If
$\delta m_i = 0$ then there is no change.

Table \ref{tab:assumptions} lists the rest of the assumptions implicit
in the basic management algorithms for memory space and CPU time.
\begin{table}[t]
  \begin{center}
    \begin{tabular}[h]{|c|l|l|}
      \hline
      \textbf{No.} & \textbf{CPU Time} & \textbf{Memory space} \\
      \hline
      1 &
      All CPU time bindings done once at start &
      All memory bindings done once at start \\
      2 &
      Time requirements constant over time &
      Space requirements constant over time \\
      3 &
      Resource requirements indivisible in time &
      Resource requirements indivisible in space \\
      4 &
      Explicit scheduling &
      Explicit allocation \\
      5 &
      Temporally independent procedures in $P$ &
      Spatially independent procedures in $P$  \\
      6 &
      $T$ is not preprocessed &
      $M$ is not preprocessed \\
      7 &
      $T$ is bound to hardware &
      $M$ is bound to hardware \\
      8 &
      $T$ is (countably) infinite &
      $M$ is (countably) infinite \\
      \hline
    \end{tabular}
  \end{center}
  \caption[Implicit assumptions]{Assumptions implicit in the symbolic
  algorithms of section \ref{sec:algorithms}.}
  \label{tab:assumptions}
\end{table}

\begin{figure}[h]
  \centering
  \epsfxsize=.5\textwidth
  \epsffile{\figdir/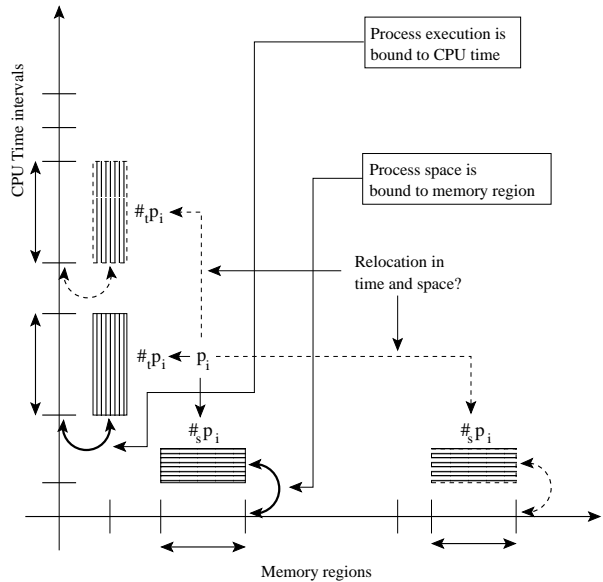}
  \caption[Relocating bindings  in time and  space]{We implicitly have
    assumed that the procedure execution and space has been bound once
    and for all  to time intervals and memory  regions.  They could be
    moved  in  the  time   and  space  yielding  spatial  or  temporal
    relocation.}
  \label{fig:moving:time:space}
\end{figure}

A procedure $p_i \in P$ is assumed to be bound to a subset of CPU time
and a subset of memory space  once and for all during its existence in
the set $P$.  It is useful  to be able to change these bindings during
the existence in  $P$; see Fig.(\ref{fig:moving:time:space}).
From  the perspective  of a  procedure $p_i$  its memory  is  bound to
memory region only once, and  its execution is bound to time intervals
only once.   We have assumed  that once established these  bindings do
not change over the procedure's life.  Relaxing this assumptions means
that $p_i$ may be unbound from  the current memory region and bound to
another memory  region over time.   Moreover its current CPU  time may
also be re-bound  to another interval over time.   This is captured in
the  first  assumption  listed  in  table  \ref{tab:assumptions}.   As
captured in the second assumption  in that table, we have also assumed
that the space and time requirements of a procedure do not change once
specified.   Our   third  assumption  is  that  the   space  and  time
requirements of a procedure must be bound in an all-or-none fashion to
the corresponding  resources.  In  other words, they  are indivisible.
The fourth assumption is  that binding the procedure time requirements
to  CPU time  and procedure  memory requirements  to memory  space are
parametrically,  i.e. explicitly, specified  by the  programmer.  The
fifth  assumption   captures  the  independence   between  procedures.
Procedures neither interact with each other in time, nor do they share
memory.   We  have also  assumed  as  our  sixth assumption  that  the
resource  sets are not  structured in  any way  through preprocessing.
Yet another assumption is that the resource sets are bound to hardware
devices.   Finally, we have  not considered  the possibility  that the
resource sets could be finite, and hence the $makeset()$ operation can
possibly fail to find a suitable subset of the resource.  Relaxing the
assumptions    implicit     in    the    expressions     of    section
\ref{sec:algorithms} offers the design space for concrete algorithms.

Memory relocation is possible when  an $Enum$ is defined over units of
memory  space and  procedure  memory requirements  can  be rebound  to
different memory spaces  during its lifetime.  Given a  finite $M$ the
$makeset()$ can fail.  If $M$ is the secondary memory then nothing can
be done.   However, if $M$ is  primary memory then some  fraction of a
secondary  memory  can  be  used  to augment  $M$,  and  then  operate
$makeset()$.  This  augmented memory cannot be used  for execution, by
definition of  primary memory.  Hence  relocation is required  to swap
regions between the primary memory and the augmented memory.  An $Org$
can be used to impose a structure on $M$.  A $Tree$ for $Org$ is often
used  over secondary  memories  to obtain  a ``directory''  structure.
Using a constant  function for $Org$ over primary  memory yields fixed
size partitioning scheme and is used to divide the primary memory into
pages or secondary  memory into blocks.  Using $\#_sp_i$  as the $Org$
yields a  variable size partitioning scheme.   Relaxing the assumption
that    $\#_sp_i$   is    indivisible   (assumption    3    in   table
\ref{tab:assumptions})  allows a division  of the  procedure's address
space using some $Org$ over  it.  A constant $Org$ divides the address
space without  any cognizance of  any structure within  the procedure.
This divides the  procedure into page sized frames  for use in primary
memory and into block sized  chunks for storage on secondary memories.
An  $Org$ that  considers  some internal  structure  of the  procedure
results in a  segmentation scheme over primary memory.   An $Org$ over
an enumerated $P$ allows  varying the $Select$ operation.  An identity
function  as  an  $Org$  over $P$  yields  the  first-come-first-serve
algorithm.   If  a sort  based  $\#_sp_i$ is  used  as  $Org$ then  an
ascending sort  yields the  shortest-size-first algorithm.  If  a sort
based on  an external $Enum$, called  $priority$, is used  we have the
priority allocation.   If the procedures are not  independent over $M$
(assumption  5 in  table \ref{tab:assumptions}),  then we  have shared
memory regions.  If the function $f_i$ in Eq.(\ref{eq:dynamic:mem}) is
specified (usually  in part) by the  programmer then the  memory is an
explicit component of the procedure $p_i$.  Otherwise the memory is an
implicit  component of  a procedure  and is  said to  be automatically
managed.  Coupled  with the $releaseset()$  allows for an  illusion of
infinite memory over finite memories.  Finally, if $M$ is not bound to
the hardware  then we  have another set  $M^\prime$, the  real memory,
that is  actually bound to  the hardware device.  $\#M^\prime$  is the
final  target of  load operations.   $M$ is  now  called \emph{virtual
  memory}, and subsets in $M$  must be bound to subsets in $M^\prime$.
Usually $\#M^\prime \le \#M$.

CPU time relocation  is possible when an $Enum$  is defined over units
of  CPU time  interval  and  procedure CPU  time  requirements can  be
rebound to  different CPU time  intervals during its  lifetime.  Since
the clock  device that  produces the  CPU time resource  is a  part of
computer hardware, the CPU time resource is continuously produced.  It
is therefore infinite, and $makeset()$  cannot fail.  There is thus no
requirement  of  techniques  analogous  to  swapping  used  in  memory
resources.  Further, since  for a single computer system  we have only
one  clock there  are no  other producers  of CPU  time.  Hence  it is
impossible  to  realize  ``time  swapping'' techniques.   The  binding
between the CPU time resource  and the procedure time requirement need
not  be  permanent  once  established.  This  allows  ``relocation  in
time'', or rescheduling.   An $Org$ can be used  to impose a structure
on $T$.   Using a  constant function for  $Org$ over primary  CPU time
yields  a part  of the  round  robin with  a fixed  time quantum.   It
identical to  framing of the memory.  Since  $\#_tp_i$ is undecidable,
using it as  the $Org$ is not possible.   Relaxing the assumption that
$\#_tp_i$ is indivisible (assumption 3 in table \ref{tab:assumptions})
allows  a division  of the  procedure's time  requirements  using some
$Org$ over it.  A constant  $Org$ divides the address interval without
any cognizance of any structure within the procedure.  This is used to
conceptually divide the procedure's execution time into chunks of same
size as the round robin quanta over $T$, and corresponds to paging the
procedure.   An $Org$ that  considers some  internal structure  of the
procedure's  time requirements  results in  ``segmentation''  over CPU
time.   A  procedure  may be  divided  into  CPU  bound or  I/O  bound
``segments''.   An $Org$  over an  enumerated $P$  allows  varying the
$Select$ operation.  An identity function  as an $Org$ over $P$ yields
the first-come-first-serve  algorithm.  If  a sort based  $\#_tp_i$ is
used  as $Org$ then  an ascending  sort yields  the shortest-job-first
algorithm.  If a multilevel queue  based on an external $Enum$, called
$priority$, is used as $Org$ we have the priority selection.  An $Org$
based  on an external  $Enum$, called  $response time  constraint$, is
used, we  have a real time  system.  Similarly, and $Org$  based on an
$Enum$  derived from events  yields event  driven scheduling.   If the
procedures  are  not  independent  over  $T$ (assumption  5  in  table
\ref{tab:assumptions}),   then  we   have   communicating  procedures.
Communicating  procedures  give  rise  to  two  interesting  problems:
synchronization of access to the shared state and deadlocks.  Although
$\#_tp_i$  is  undecidable,  it  is  possible  to  define  a  function
analogous  to Eq.(\ref{eq:dynamic:mem}) that  can attempt  to estimate
the future CPU  time requirements, or even arbitrarily  define it.  If
it is  specified by the  programmer then the  CPU time is  an explicit
component  of the  procedure  $p_i$.   Otherwise the  CPU  time is  an
implicit  component of  a procedure  and is  said to  be automatically
managed.   CPU  time  is  often  automatically  managed  in  practice.
Estimates of  required CPU  time are useful  in real time  systems for
decisions like load shedding.   Arbitrary specification of CPU time is
used  to limit resource  consumption or  to obtain  specific execution
times.  Finally,  if $T$  is not  bound to the  hardware then  we have
another set $T^\prime$,  the real CPU time, that  is actually bound to
the  hardware  device.   $\#T^\prime$  is  the  final  target  of  run
operations.  $T$ is now called \emph{virtual CPU time}, and subsets in
$T$ must be bound to subsets in $T^\prime$.

\subsection{A concrete algorithm for conventional OSes}
\label{sec:concrete:algos}

The assumptions  implicit in the abstract algorithms  offer the design
space  of concrete algorithms  of an  OS.  I  illustrate this  for one
typical algorithm of an OS:  the paging and segmentation based primary
memory   management.    The  resulting   algorithm   is  in   appendix
\ref{sec:mm:with:page:segment}.     In   this   section    I   discuss
instantiating the abstract algorithms  using the assumptions to obtain
an algorithm nearer to practice.  Low level details are skipped in the
interest of clarity, and  hence only the essential abstract operations
are instantiated.   Thus while the $Org$ operation  is instantiated to
organize the  primary memory, binding  a program to primary  memory is
essentially  a  load  operation  at  low enough  levels  and  is  left
uninstantiated.

\subsubsection{Paging and segmentation}
\label{sec:concrete:paging:seg}

\begin{figure}[ht]
  \centering
  \epsfxsize=.7\textwidth
  \epsffile{\figdir/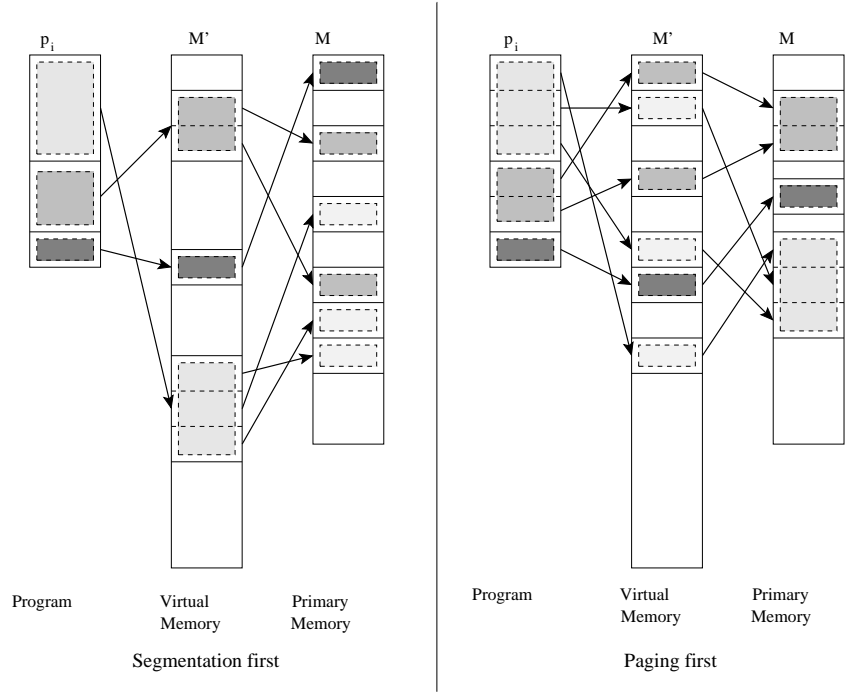}
  \caption[Paging   and   Segmentation]{The   first   of   paging   or
    segmentation of a program $p_i$ uses virtual memory $M^\prime$ and
    the  second uses  the  physical memory  $M$.   The virtual  memory
    $M^\prime$ is larger than  the physical memory $M$.  Three regions
    of a program $p_i$ are eventually mapped to the primary memory.}
  \label{fig:paging:segmentation}
\end{figure}

Using both  segmentation and  memory requires introducing  the virtual
memory $M^\prime$ as the target of  the first of the two, and then use
the second to map the virtual memory $M^\prime$ to the physical memory
$M$,  as   shown  in  Fig.(\ref{fig:paging:segmentation}).    $M$  and
$M^\prime$ use  the set of  natural numbers, $\mathbb{N}$,  to address
memory elements,  i.e. $Enum: M, \mathbb{N} \rightarrow  M$ and $Enum:
M^\prime,  \mathbb{N}  \rightarrow  M^\prime$  are defined.   In  what
follows  we assume that  all other  assumptions are  maintained.  This
prevents  us   from  introducing  techniques  like   page  or  segment
relocation  or demand  loading or  changing the  memory  allocation at
runtime.

A  program must  be divided  into segments  before the  memory  can be
organized  since  the  segment  information  is  required  for  memory
organization.  Paging is independent  of the program itself, and hence
only needs the page size parameter.  The principle of binding requires
that bindings  of program  division be computed  before the  memory is
structured.  Virtual  memory must be introduced to  employ both paging
and segmentation, with  the first of the two  using the virtual memory
as the  target.  For concreteness  let us assume that  segmentation is
employed first followed by  paging.  Thus the segmentation information
must  be used  to  compute the  virtual  memory organization.   Paging
calculations consist of organizing physical memory as page frames, and
mapping virtual  memory segments to page  frames.  Frame organizations
are independent of all others and may be performed at any earlier time
as permitted by the  binding principle.  Segmenting the virtual memory
and  framing the  physical  memory are  independent  since the  target
resource  sets  are  different   for  each.   Segmenting  and  framing
computations can  be ordered  in any sequence  in accordance  with the
binding principle.   Mapping operations must  be the last step  of our
algorithm.  Page  replacement strategies need not  be considered since
sufficient  free  memory  is   assumed  to  be  available  and  memory
requirements do not change during execution.

One sequence in accordance with  the binding principle consists of the
following steps.
\label{paging:algo:schematic}
\begin{enumerate}
\item Segment the program.
\item Divide  the virtual  memory $M^\prime$  into segments  using the
  segmenting information.
\item Build the  segment table that binds program  segments to virtual
  memory segments.
\item Divide  the virtual  memory segments into  pages using  the page
  size parameter.
\item Frame  the physical memory $M$ using the page size parameter.
\item Build  the page  table that  binds the  virtual memory  pages to
  physical memory frames.
\end{enumerate}
The  sequencing arguments are  similar to  hoisting arguments  used in
program analysis  and are  based on the  binding principle.   The step
that organizes the  primary memory into frames is  the $Org$ operation
over $M$,  and can be  hoisted to any  place before the page  table is
built.  The first  step, segmenting the program, is  often bound early
during compilation stage and cached.   Hence when a procedure is to be
bound  to primary memory  the OS  can read  off the  cached segmenting
information and quickly move to the next step.

Given the concrete $Enum$  and $Org$ operations, the memory management
predicate   (Eq.(\ref{eq:basic:mgmt}))   can   be   instantiated   for
$M^\prime$     and     $M$.       The     memory     invariants     in
Eqs.(\ref{eq:total:size:invariance})-(\ref{eq:disjointness})       are
preserved by Eq.(\ref{eq:basic:mgmt}).

The  paging and  segmentation algorithm  is symbolically  expressed in
\ref{sec:mm:with:page:segment}.

\subsection{Summary}
\label{sec:single:cpu:summary}

Our definitions have served to define the goal of an OS.  The matching
principle has given a structure in Fig.(\ref{fig:os:abstractions}) for
a  conventional  OS.   Section  \ref{sec:algorithms}  shows  that  the
equivalence  class partitioning  of  resources can  realize the  basic
constructs  to   support  the  high  level  machine   as  defined  for
conventional OSes.  It employs the  binding principle in two ways.  It
also exposes the  common structure within the algorithms  for CPU time
and     memory      space     manipulations.      Finally,     section
\ref{sec:algorithms:concrete} examines the implicit assumptions in the
symbolic expressions of the algorithms and reveals the design space as
well as further opportunities to employ the binding principle.

\section{A sketch of the formal structure of an OS}
\label{sec:os:theory:sketch}

A clear definition  of an OS, some principles  involved in its design,
and a symbolic as well as  concrete description of algorithms of an OS
enable a high level view that can be used to devise a formal structure
of an OS.  In this section I attempt to sketch such a structure.

Our definitions open a simple  description of a formal structure of an
OS: a process algebra of  a universal machine.  A universal machine is
given a set  of procedures for interpretation and  its task is defined
to  effect their  interpretation  as specified.   The first  classness
principle must be employed to switch between active and passive states
of   the  procedures.   

Let $P$ be a set of procedures  $p_i$, $i \in \mathbb{N},\ 0 \le i \le
N$ and $E$ denote the interpretation context, i.e. the environment.  A
change in the system state changes $E$.  A closure is a subset of $E$.
The procedure $p_0$  is the ``null'' procedure.  It  occupies the free
memory space, the  free CPU time and does not change  the state of the
system.  Three  operators, $close: p  \rightarrow E$, $schedule:  E, p
\rightarrow  E$  and  $resume:  E  \rightarrow  p$  are  required  for
switching.   The $close$  operator  computes the  closure  of a  given
procedure and  stores it.  The $schedule$ operator  returns the stored
closure  of the  next procedure  to schedule.   The  $resume$ operator
restores the  closure given by  the $schedule$ operator.   Let ``$;$''
denote the  sequence operator in  infix form so  that $a\ ;\  b$ means
that execution of  $b$ immediately follows the execution  of $a$.  Let
``$<$'' denote the ``before'' relationship  in infix form so that $a <
b$ means that  $a$ executes before $b$.  The  $close$ operator and the
$schedule$  operator always  precede the  $resume$ operator.   Thus we
have
\begin{enumerate}
\item $close\ <\ resume$
\item $schedule\ <\ resume$
\end{enumerate}
as  two  universal  invariants  of these  operators.   The  $schedule$
operator may commute with the  $close$ operator over the $;$ operation
for general purpose  OSes but not for specialized  OSes like real time
OSes.  Its commutativity over $;$ is therefore not universal.

The $switch: P, \mathbb{N} \rightarrow p$ operator can be defined as:
\begin{equation}
  \label{eq:switch:op}
  switch \names Sel\ ;\ close\ ;\ Sel\ ;\ schedule\ ;\ resume
\end{equation}
and is the  continuation that an OS uses  to execute procedures almost
as coroutines.   The first $Sel$  operator selects the  current active
procedure  $p_i$  that is  to  be  deactivated  and the  second  $Sel$
operator selects  the next  procedure $p_j$ that  is to  be activated.
The $schedule$  operator looks up  the environment $E$ to  extract the
stored closure of $p_j$.  If the first $Sel$ and the $close$ operators
are part of the procedure $p_i$ then the OS is said to $cooperatively\
multitask$, and  the procedures execute exactly as  coroutines.  An OS
becomes $idle$ if $p_j = p_0\  \wedge\ p_i \not = p_0$.  An OS becomes
$loaded$  if  $p_j  \not =  p_0\  \wedge\  p_i  = p_0$.   The  binding
principle  must  place the  execution  of  $close$  and $schedule$  in
accordance  with the  two invariants  above.  

More specifications  are needed for a  complete formal view  of an OS.
For instance  the invariants in the behavior  of interacting processes
must  be captured from  an OS's  point of  view.  The  two fundamental
procedure state transitions  are ``running'' and ``ready-to-run'', and
the well known state transitions like waiting, ready to run, sleeping,
swapped etc. must also be  formally captured as specific variations of
these  two.  Another  invariant that  a formal  theory of  an  OS must
capture is the orthogonal execution property of an OS, i.e. an OS must
not change the semantics of individual procedures.  One way to do that
is to associate  a context to each procedure $p_i \in  P$ with the OS'
own context  associated with  the null procedure  $p_0$.  Thus  we may
demand that  the orthogonal  execution property induce  an equivalence
class partitioning of the complete environment of the computer system.
Finally,  the  equivalence class  partitioning  approach implies  that
protection schemes  are simply guards over each  subset.  These guards
are  a  part  of  the  rules that  capture  the  orthogonal  execution
property.

Conventional  OSes are  seen as  a  specific instance  of the  general
definition of an OS and include the general structure in the preceding
paragraphs.  For such OSes the  first classness principle allows us to
treat the CPU time and memory space resources uniformly.  It allows us
to  recast the  resource management  problem in  terms  of equivalence
class partitioning of  an infinite countable set that  denotes a given
resource.  In  particular the structural similarity  of algorithms for
different resources emerges from  their symbolic forms, and the design
space  for concrete algorithms  is easily  extracted by  examining the
entities in them.  The design  space, captured as a set of assumptions
in section  \ref{sec:algorithms:concrete:assumptions}, can be formally
captured by defining suitable functions like Eq.(\ref{eq:dynamic:mem})
and      identifying     the      general      invariants.      Unlike
Eq.(\ref{eq:dynamic:mem}) they can be resource agnostic.

An effort towards the formal structure of an OS needs substantial care
and will be undertaken in future.

\section{Related work}
\label{sec:lit:review}

Historically, the attempts  of the ``How?''  of OS  have served as the
basis        of        defining        an        OS.         Solntseff
\cite{Solntseff:1981:SOS:1164685.1164689}   succinctly   captures  the
evolution  in terms  of era  like the  initial ``topsy''  period where
everything not  directly produced by  the user was considered  part of
the system, the  ``compatibilist'' view that saw OS  as a mechanism to
provide machine  independence, the ``perfectionist'' view  that saw OS
as   a  mechanism  to   correct  hardware   design  errors,   and  the
``allocationist''  period  that saw  an  OS  as  a manager  of  finite
resources.      In      an     early     review      paper     Denning
\cite{Denning:1971:TGC:356593.356595} offered a definition of an OS in
terms  its seven  supervisory and  control functions.   An interesting
footnote  in  the  paper  observed   that  a  process  is  a  ``direct
generalization'' of procedure  in execution, but did not  build on it.
The  review went  on  to identify  five  abstractions --  programming,
storage  allocation,  concurrent  processes, resource  allocation  and
protection -- that could form the  basis of a ``theory'' of OS.  While
a lot of work has been  done since then, the core ideas and algorithms
discussed in this paper have largely remained unchanged.  Denning also
reviewed  the  definitions  of  a  process  and  noted  that  although
imprecise, they  were sufficient for  implementation purposes.  Dennis
and  van Horn's  \cite{Dennis:1966:PSM:365230.365252}  was an  earlier
attempt at figuring out  the semantics of multiprogrammed systems.  As
noted      in      a     later      reprint      of     the      paper
\cite{Dennis:1983:PSM:357980.357993}, it used different terminology to
define an OS as an  high level abstract machine using data abstraction
concepts.

The creative  and innovative period  of OS research probably  were the
1960 and  1970 decades when most  of the ideas in  practice today were
conceived, investigated and developed.  The process as an abstract and
the  central  entity  within  an  OS  was  recognized.   Most  of  the
metatheory  was developed  for specific  sub problems  within  the OS.
Processes  were  defined  in  various  ways,  some  indirectly.   Holt
\cite{Holt:1972:DPC:850614.850627}  viewed a  process as  an ``agent''
that          causes         state          changes.          Dijkstra
\cite{Dijkstra:1968:SLS:363095.363143}   saw  it  as   a  ``sequential
automaton''.  Dennis and van Horn \cite{Dennis:1966:PSM:365230.365252}
described   it    as   a    ``locus   of   control'',    and   Denning
\cite{Denning:1971:TGC:356593.356595}  presented   the  ``instance  of
program  in  execution'' view,  and  reviewed others.   Implementation
techniques for managing memory,  processes and devices were developed.
See  \cite{Denning:1971:TGC:356593.356595}   for  a  review.   Viewing
processes as formal computation subsumes these varied views.

The late 70s and early 80s  saw the birth of personal computing, and a
surge in innovations in hardware technologies.  A significant part was
the  hardware support  for  OS  operations, which  in  turn drove  the
systems  research.  The trend  continues to  this day.   Projects like
Hydra            \cite{Wulf:1975:OHO:1067629.806530},           Amoeba
\cite{Tanenbaum:1990:EAD:96267.96281},                           Medusa
\cite{Ousterhout:1980:MED:358818.358823},                        Spring
\cite{Stankovic:1989:SKN:71021.71024}            and            Accent
\cite{Rashid:1981:ACO:800216.806593}    focused   on    the   evolving
challenges  of multiprocessor  and distributed  systems.   Others have
focused  on individual sub  problems in  such systems.   For instance,
issues   like   programmability   \cite{Marques:1989:EOS:74877.74890},
monitoring and  debugging \cite{Tokuda:1988:RMD:68210.69222}, parallel
scheduling    \cite{Narang:2011:PDM:2007183.2007186}   and   multicore
scheduling    \cite{Shelepov:2009:HSH:1531793.1531804}    have    been
addressed               in               the               literature.
Distributed systems are challenging and it is
important to identify specific problems  and issues.  We offer a clear
view of processes in a distributed system.

Architectural issues in OS have also been worked on.  Apertos (earlier
Muse)
\cite{Yokote:1992:ARO:141936.141970,Yokote:1991:MOA:122120.122122,Yokote:1992:NSO:506378.506427}
attempts to  bring in reflectivity \cite{Smith:1984:RSL:800017.800513}
primarily motivated by the advent of mobile computing.  The separation
of  object level and  meta level  abstractions represented  within the
same framework is  the central thesis of this  work.  N\"urnberg \etal
\cite{Nurnberg:1996:HOS:234828.234847} elevate  the view of Hypermedia
from a paradigm  of information organization to a  view of a computing
paradigm.   The  separation of  data,  structure  and  behavior is  an
interesting idea  in this work.  Refining  the ``structure'' component
with ``application''  of the  $\lambda$ calculus, their  insight could
connect well with the definition  of a conventional OS offered in this
work.  Factored  OS (FOS) \cite{Wentzlaff:2009:FOS:1531793.1531805} is
an approach to deal with the growing complexity of OS.  It is based on
factoring a component (service) into  smaller ones and is motivated by
the need for OSes to scale up for multicore systems.  This corresponds
well with the use of the binding principle to guide the factorization.
In  contrast to  these design  time  variations of  the OS  structure,
approaches like Exokernels or  SPIN investigate techniques of mutating
the    structure     at    runtime    of     the    OS.     Exokernels
\cite{Engler:1995:EOS:224056.224076} are the result of re-architecting
traditional  OS  structure  to  safely expose  physical  resources  to
applications to allow application  specific customization.  This is an
example of redefining the interface,  or the abstract machine, that an
OS presents  to applications.   Viewing from a  functional programming
perspective, exokernels  could be the design choice  that, in general,
has  an  OS return  carefully  packaged  procedures  for safely  using
physical  resources.   The SPIN  \cite{Bershad:1995:ESP:224057.224077}
approach  tries  to  employ   the  good  properties  of  the  Modula-3
programming language to obtain an extensible system from a core set of
extensible  services.  This safely  changes the  interface that  an OS
presents  to  an  application,  and  is an  example  of  the  matching
principle used to identify the core set of services.  While individual
issues to  focus appear distinct,  the common underlying  concerns are
about the ``right'' approach to  structuring, if any.  Our work offers
the binding principle as a concrete approach to structuring.

The growing  complexity of  the OS problem  has greatly  increased the
turnaround time for building experimental systems and efforts to build
such systems have  reduced.  The decade long K42  effort by Wisniewski
\etal   \cite{Wisniewski:2008:KLO:1341312.1341316}    is   a   notable
exception. They  explore the facets of  building a full  OS, share the
experience and  insights, and point  out to the  need to fill  the gap
between good  values of research (producing  meaningful results beyond
microbenchmarks) and  actual research practice (high  cost of complete
OS research against time-cost  constraints).  They point out that some
important  practical  questions,  e.g.   the useful  lifetime  of  the
GNU/Linux or  Windows structures, are unanswered,  and offer plausible
reasons for the lack  of whole-OS research efforts.  Whole-OS efforts,
e.g. to  describe the structure  of an OS,  are needed in  addition to
incremental work.

Another approach to deal with  complexity is to ``go small'', and some
efforts  have  investigated  OS  on  single user  systems.   Stoy  and
Stratchey \cite{stoy:strachey:os6:1,stoy:strachey:os6:2}, for example,
set  out to  develop the  OS6 which  was an  early attempt  at virtual
machines  and avoided a  job control  language through  a hierarchical
control structure for system  use (as opposed to hierarchical resource
allocation).  Single  user OSes  also permitted removing  the boundary
between       the      OS       and       the      user       program.
\cite{Ranai:1986:DRS:382158.383030}  Such  attempts  could  have  been
useful to extract the  essential abstractions, but further work needed
has not been pursued.

On the formal  side, attempts have been made to build  models of OS --
in  whole or in  part, develop  languages to  express them  and verify
them.  Yates \etal \cite{Yates99i/oautomaton} developed a formal model
of an OS as a  system of distributed state machines. They investigated
two views of an OS: a  user level model as an interface specification,
and  a kernel  level  model  of the  implementation  that exposes  the
details  hidden by  the user  level abstraction.   They show  that the
kernel level model  indeed implements the user level  model, and hence
both  are functionally equivalent.   Another example  of the  value of
formal work in OS, or its parts, is the graph theoretic description of
the deadlocks  problem \cite{Holt:1972:DPC:356603.356607}.  It brought
a number of previous results together into a simple neat structure and
lead to  efficient deadlock  detection and prevention  algorithms.  On
the         verification          front,         Barreto         \etal
\cite{Barreto:2011:ASF:1945023.1945042}  specified process management,
IPC and file system components of  an OS kernel in Z, uncovered errors
and inconsistencies in the kernel  and formally verified it by using a
mechanical theorem prover  Z-EVES on the Z specification.   We can see
the current work  as yielding a framework for  formal development, and
is hence semiquantitative in nature.

Ideas  from  programming  languages  have  been  used  to  investigate
structure     and     structuring     issues     in     OS.      Clark
\cite{Clark:1985:SSU:323627.323645}  discussed the  use of  upcalls to
structure       programs       like       an       OS.        Kosinski
\cite{Kosinski:1973:DFL:390014.808289}   identified   eight  important
issues in  expressing OS code,  e.g.  the need for  parallel operation
yet be  determinate, or  understandability (of OS  code) in  the large
etc.  He developed  a data flow language based  on function definition
and composition, tried to identify minimal computational function, and
used these to sequence  computations.  The Barrelfish effort developed
the  Filet-o-Fish  language \cite{Dagand:2010:FPD:1713254.1713263}  to
construct  a  domain  specific  languages  (DSL) and  employed  it  to
generate low level OS code.  Back \etal \cite{Back98javaoperating} use
Java to explore  the OS design space, and  outline the major technical
challenges.  They  point out that adapting language  technology to fit
into  OS framework  could be  used to  deal with  the  challenges.  In
particular, they show how garbage collection techniques can be used to
support        resource         management.         Flatt        \etal
\cite{Flatt:1999:PLO:317765.317793} demonstrate  how key OS facilities
are obtained through three  key extensions -- threads with parameters,
eventspaces, and  custodians -- to a high  level programming language.
They  summarize an important  lesson in  the title  of their  paper --
``Programming   languages  as   operating  systems''.    The  matching
principle in the current work is useful to understand their work.  The
current work  attempts to concretely  use these insights  by exploring
the  consequences of  a  focus  on program  execution  as the  central
concept at which programming languages and OSes converge.

In summary, the  initial period of work dealt with  the ``How?'' of an
OS, and most of the algorithms  in use today have been investigated in
this period.   While the growing  complexity aggrandized the  focus on
the ``How?'',  some approaches to  see the structure within  have also
emerged.   Various  directions  of  attack --  ``go  small'',  explore
alternate structuring techniques,  borrow ideas from other disciplines
like programming  languages etc.  have  been explored.  Investigations
into the formal structure are  needed to achieve generality as opposed
to specific case  solutions that have been the  main ``How?'' response
to fast  evolving challenges.  Experimental whole-OS  efforts like the
K42  are  a  necessary  complement  to formal  efforts  like  the  I/O
automaton model or  specification in Z.  A good  model of OS structure
that identifies  the orthogonal concepts  and uses them to  devise the
specific  algorithms is  also  pedagogically useful  and  can help  in
activities like  OS course design \cite{Creak:2000:TOS:506117.506126}.
For  instance, traditional  memory  management and  file systems  have
redundancies  that  can  be  eliminated  to  effectively  teach  those
algorithms \cite{Esser11}.

\section{Conclusions}
\label{sec:conclusions}

The rapid  pace of  systems evolution has  resulted in a  very systems
oriented work  at the expense of  attention to the ``what  is an OS?''
question.   Steps towards  answering that  question  could potentially
help to manage  the complexity by offering at  least some organization
of the know-how  that has been so meticulously  developed.  This paper
takes some steps towards the ``what is an OS?'' question.  It offers a
descriptive and a  constructive definition of an OS  based on concepts
from  theoretical computer science  and programming  languages.  Ideas
from  the programming  languages work  are used  to offer  some useful
principles  to  guide  the   development  of  the  algorithms.   These
definitions  and   principles  must  encompass   the  conventional  OS
practice.   We  therefore  specialize  the general  OS  definition  to
capture practice  and employ the principles  to symbolically construct
the algorithms.   The symbolic  expression is based  on the  idea that
resource  subsets must  induce  an equivalence  class partitioning  to
realize a  universal machine  that starts  with a set  of one  or more
procedures  (Eq.(\ref{eq:basic:set:of:files})).  This  construction is
facilitated by identifying  the essential abstract operations required
by  an OS.   This  brings  out the  essential  similarity between  the
various algorithms.   In particular, we  see that the  differences are
rooted in  the properties  of the resources  over which  they operate.
The symbolic algorithms are  critically analyzed to explore the design
space  they offer.  A  particular algorithm  in practice  is extracted
that connects well with practice.

This  work attempts  to unify  the  various attempts  into a  coherent
framework  by  an  explicit  definition  of an  OS,  identifying  some
essential principles, and using these to build a generic framework for
conventional OSes.   The result connects  well with some of  the major
directions  in which  the  OS  problem has  been  addressed.  It  also
explicitly demonstrates the uniformity  of the internal algorithms via
partial  evaluation arguments  and lends  support to  the observations
like in  \cite{Esser11}.  It offers  insight into some  ingredients of
the formal  structure, and  suggests techniques from  program analysis
domain  that might  be useful  for further  work.  Efforts  like Flatt
\etal  \cite{Flatt:1999:PLO:317765.317793}   show  that  there   is  a
connection  between  programming  languages  and OSes.   Although  our
algorithms  are precise  symbolic  expressions, a  deeper study  would
investigate the rules of instantiation and composition of the abstract
operations.  In  essence, an OS could  be regarded as  an algorithm to
obtain a  desired model of  computation over another given  model.  It
thus forms a constructive proof of equivalence between the computation
models when Turing complete computation  models are chosen as the high
and low level machines of an OS.

\subsection{Future work}
\label{sec:future}

The current work sketches a formal structure of the solution of the OS
problem by clearly  defining an OS, and devising  the framework of the
solution.  This opens up a number of possibilities and questions, both
theoretical  and practical.   First and  perhaps the  most interesting
aspect  is that  a formal  completion of  this approach  results  in a
detailed algorithm to convert  an imperative machine into a completely
functional one  for conventional OSes.   While we argue that  the high
level functional  machine is a  good target machine  for an OS,  it is
easy to see  that an OS is essentially an algorithm  that allows us to
move from one model of  computation to another.  Second, the framework
resulting  from this work  helps to  better identify  parts of  the OS
problem in use today that are not yet general enough.  It can show the
prospective paths to pursue.  A functional style of expressing OS code
could  be organize  the components  in a  much generic  way  given the
insights of  the framework.   Third, it suggests  that the  problem of
distributed operating systems could be reduced to an equivalence proof
of formal  computation model  of such systems  to a well  known model.
Fourth,   the   arrangement   of   levels   of   abstraction   as   in
Fig.(\ref{fig:prog:lang:abstractions})                             and
Fig.(\ref{fig:os:abstractions}) is only intuitive.  Can the concept be
formalized?  Fifth,  a rather interesting question  would be regarding
the nature  of objects  given their first  classness so that  parts of
their  lifetimes are  as dynamic  entities, processes,  and  as static
entities, files,  otherwise.  Could first classness be  used to devise
some variational  principle that encompasses the  variety of execution
possibilities  relevant for program  expression and  program execution
support?

A mathematical theory of OS would specify properties of the operations
of  section  \ref{sec:universal:os:functions}  like whether  they  are
commutative, under what conditions etc.  The definitions of OS offered
in this  work suggest an explicit  focus on the  execution of computer
programs,  while  programming  languages  focus  explicitly  on  their
expression aspects  and implicitly focus on their  execution.  The key
insight  is  that  the  execution  model,  the  high  level  model  of
computation, is  common to  program expression and  program execution.
There  is a  significant body  of  work in  the programming  languages
community  that could  be useful  to model  OS.  Techniques  like flow
logic  for   process  calculi,  abstract  state   machines,  or  sound
approximations through abstract interpretation  could be brought in to
study  the structure  of OS.   A mathematical  theory of  an  OS would
essentially be  a process algebra of  processes satisfying definitions
like in section \ref{sec:defns}.

\section*{Acknowledgments}
\label{ack}
I have  benefited from discussions  with my earlier  colleagues Rustom
Mody and Achyut  Roy at University of Pune, and  with Uday Khedker and
Amitabha Sanyal during  my stay at I.I.T.  Bombay.   Thanks to my many
students who have borne me through the development of these ideas.

\bibliographystyle{unsrt}
\bibliography{refs}

\appendix

\section*{APPENDIX}
\section{Paging and segmentation based memory management}
\label{sec:mm:with:page:segment}

This section  illustrates the details  of the paging  and segmentation
based  memory  management  technique.   We ignore  the  implementation
details  like the  availability of  pointers and  dedicated registers.
For concreteness we segment first and then page the procedures.

The schematic  of the  algorithm for paging  and segmentation  on page
\pageref{paging:algo:schematic} has  three sets over  which operations
must  be performed.   They are:  (a) the  set of  numbers  forming the
procedure, (b) the set of virtual  memory elements, and (b) the set of
physical memory elements.  Let the numbers $n_i$ in a procedure $p_i$
be denoted as:
\begin{equation}
  \label{eq:denoting:nums:in:procs}
  p_i \equiv \{n_{i_1}, n_{i_2}, n_{i_3}, \ldots\},
\end{equation}
and let $n_i \equiv \#p_i$, denote the size of the set $p_i$.  I first
specialize the  $makeset$ of Eq.(\ref{eq:subset:function}).  $makeset$
aims at  obtaining a  set of  given size from  within anywhere  in the
resource.   In contrast,  the  $makespecset$ returns  the subset  from
within the specific part of the  resource.  Its use must be guarded by
the invariance demands on page \pageref{eq:total:size:invariance}.
\begin{eqnarray}
  \label{eq:specific:subset:function} 
  &~&
  makespecset (S,\ u_{min},\ u_{max}) 
  \nonumber \\
  &~&
  \quad\quad
  \names\ \biggl(\forall j \in \mathbb{N}\ \vert\  (u_{min} \le j \le
  u_{max})\ :\ 
  Sel(Enum(S,\mathbb{N}), j)
  \biggr)\biggr)
\end{eqnarray}

Segmenting a  given procedure $p$  accepts the number of  segments and
the  segment information  as  parameters, and  uses the  $makespecset$
(Eq.(\ref{eq:specific:subset:function})) operation to build its set of
segments.   Paging a  given segment  $s$ accepts  the page  size  as a
parameter, and  uses the $makespecset$  operation to build  its pages.
Segmenting the  virtual memory $M^\prime$ accepts the  set of segments
of a  procedure as  a parameter, and  uses the $makeset$  operation to
carve the  segments in the virtual  memory.  The next  operation is to
bind  the segments of  the procedure  to the  segments in  the virtual
memory.   Conceptually, this  loads the  procedure into  the segmented
virtual memory.  Paging the physical  memory $M$ accepts the page size
as a parameter, and uses the $makeset$ operation to frame the physical
memory.  Finally, the pages of  the bound (i.e.  loaded) segments of a
procedure are bound to the frames.  Since the memory is physical, this
corresponds  to  loading  the  procedure pages  into  physical  memory
frames, and completes the algorithm.

The  invariants  on  page \pageref{eq:total:size:invariance}  must  be
defined for  each of the  two memories, the virtual  memory $M^\prime$
and   the    physical   memory   $M$.     Let   $S_O^{M^\prime}$   and
$S_F^{M^\prime}$  denote the  sets  of occupied  and  free regions  of
$M^\prime$.  Let  $S_O^{M}$ and $S_F^{M}$ denote the  sets of occupied
and free regions of $M$.

The  symbolic expressions for  the algorithm  that first  segments and
then pages procedures while loading onto physical memory are:

\begin{enumerate}
\item   The  $procSegs$  algorithm   in  Eq.(\ref{eq:proc:segmenting})
  segments  a procedure  $p$ given  its  number of  segments $k$,  and
  segment start offset addresses $a_1 = 0, a_2, \ldots, a_k, a_{(k+1)}
  = \#p$,  
  \begin{eqnarray}
    \label{eq:proc:segmenting}
    &~&
    procSegs(p, k, a_1, a_2, \ldots, a_k, a_{k+1}) 
    \nonumber \\
    &~& \quad \quad \names \biggl(
    \forall i \in \mathbb{N}\ \vert\ (1 \le i \le k)\ :\ 
    makespecset(p, (a_{i+1} - a_i - 1))
    \bigr)\biggr)
  \end{eqnarray}
\item  The  $makeSegs$  algorithm in  Eq(\ref{eq:proc:segment:mprime})
  divides  the  virtual  memory  $M^\prime$ into  segments  using  the
  segmenting information.   The $makeset$ operation is  used since the
  segment  can be  carved from  anywhere  in the  virtual memory.   It
  yields  the set  $\{R_i\}$ of  segments in  the virtual  memory, and
  preserves the invariances for virtual memory.
  \begin{eqnarray}
    \label{eq:proc:segment:mprime}
    &~&
    makeSegs(M^\prime, k, a_1, a_2, \ldots, a_k, a_{k+1}) 
    \nonumber \\
    &~& \quad
    \names \biggl(
    \forall i \in \mathbb{N}\ \vert\ (1 \le i \le k)\ :\ 
    \nonumber \\
    &~& \quad \quad  \quad
    \bigl(\mathrm{Let}\ (r_i \names (a_{i+1} - a_i - 1)) \wedge
    (\#S_F^{M^\prime} \ge r_i)\ \mathrm{in} 
    \nonumber \\
    &~& \quad \quad \quad \quad  \quad(\mathrm{Let}\ (R_i \names
    makeset(S_F^{M^\prime}, r_i)) \ \mathrm{in}  
    \nonumber \\
    &~& \quad \quad \quad \quad \quad \quad  \quad
    remove (S_F^{M^\prime}, R_i) \wedge
    append(S_O^{M^\prime}, R_i)
    )\bigr)\biggr)
  \end{eqnarray}
\item  The   $segTab$  algorithm  in  Eq.(\ref{eq:proc:build:seg:tab})
  builds  the segment  table that  binds program  segments  to virtual
  memory segments.
  \begin{eqnarray}
    \label{eq:proc:build:seg:tab}
    &~&
    segTab (M^\prime, p, k, a_1, a_2, \ldots, a_k, a_{k+1})
    \nonumber \\
    &~& \quad \quad  
    \names \biggl(
    \mathrm{Let}\ (ps\ \names\ \left\{procSegs(p, k,
      a_1, a_2, \ldots, a_k, a_{k+1})\right\})\ \mathrm{in}
    \nonumber \\
    &~& \quad \quad \quad \quad
    \bigl(\mathrm{Let}\ (s\ \names\ \left\{makeSegs(M^\prime,
    k, a_1, a_2, \ldots, a_k, a_{k+1})\right\})\ \mathrm{in}
    \nonumber \\
    &~& \quad \quad \quad \quad \quad \quad
    \bigl(\forall i \in \mathbb{N}\ \vert\ (1 \le i \le k)\ :\  
    bind (Sel(ps, i), Sel(s, i)) 
    \bigr)\bigr)\biggr)
  \end{eqnarray}
  Given    the    set   of    segments    of    the   procedure    $p$
  (Eq.\ref{eq:proc:segmenting}),  and  the  set  of  segments  in  the
  virtual           memory          (Eq.\ref{eq:proc:segment:mprime}),
  Eq.(\ref{eq:proc:build:seg:tab})  selects the $i^{th}$  segment from
  each and binds them together.

\item  The   $makePage$  algorithm  in  Eq.(\ref{eq:proc:page:mprime})
  divides  a virtual  memory segment  into pages  using the  page size
  parameter $g$.

  \begin{eqnarray}
    \label{eq:proc:page:mprime}
    &~&
    makePage(s, g)
    \nonumber \\
    &~& \quad
    \names\ \biggl(
    \mathrm{Let}\ (\#g \names \lceil \#s/g \rceil)\ \mathrm{in}\
    \nonumber \\
    &~& \quad \quad \quad \quad 
    \bigl(\forall i \in \mathbb{N}\ \vert\ (1 \le i \le \#g)\ :\  
    \nonumber \\
    &~& \quad \quad \quad \quad \quad
    makespecset(s, ((i-1) \star \#g), (i \star \#g))
    \bigr)\biggr)
  \end{eqnarray}
\item The  $makeFrame$ algorithm in  Eq.(\ref{eq:proc:frame:m}) frames
  the physical memory $M$ using  the page size parameter.  It yields a
  set $\{fr_i\}$  of frames and  preserves the invariances  for physical
  memory.
  \begin{eqnarray}
    \label{eq:proc:frame:m}
    &~&
    makeFrame(M, g)
    \nonumber \\
    &~& \quad
    \names\ \biggl(
    \forall i \in \mathbb{N}\ \vert\ (1 \le i \le \#g)\ \wedge\
    (\#S_F^M \ge \#g)\ :\  
    \nonumber \\
    &~& \quad \quad \quad \quad \quad
    (\mathrm{Let}\ (fr_i \names makeset(S_F^M, g))\ \mathrm{in}
    \nonumber \\
    &~& \quad \quad \quad \quad \quad \quad
    remove (S_F^M, fr_i) \wedge append(S_O^M, fr_i)
    \bigr)\biggr)
  \end{eqnarray}

\item  The  $pageTab$  algorithm in  Eq.(\ref{eq:proc:build:page:tab})
  builds  the  page table  that  binds  the  virtual memory  pages  to
  physical memory  frames.
  \begin{eqnarray}
    \label{eq:proc:build:page:tab}
    &~&
    pageTab (M, s, g)
    \nonumber \\
    &~& \quad \quad  
    \names \biggl(
    \mathrm{Let}\ (pgSet\ \names\ \left\{makePage(s, g)\right\})\
    \mathrm{in} 
    \nonumber \\
    &~& \quad \quad \quad \quad
    \bigl(\mathrm{Let}\ (frSet\ \names\ \left\{makeFrame(M,
      g)\right\})\ \mathrm{in} 
    \nonumber \\
    &~& \quad \quad \quad \quad \quad \quad
    \bigl(\forall i \in \mathbb{N}\ \vert\ (1 \le i \le \lceil \#s/g
    \rceil)\ :
    \nonumber \\
    &~& \quad \quad \quad \quad \quad \quad \quad \quad
    bind (Sel(pgSet, i), Sel(frSet, i)) 
    \bigr)\bigr)\biggr)
  \end{eqnarray}
\end{enumerate}

We  can now  put all  of  the above  together into  an algorithm  that
segments over virtual memory $M^\prime$ and pages over physical memory
$M$  for  each  procedure  in  $P$.  For  convenience,  we  assume  an
operation  -- $getSegsData$ that  returns the  number of  segments and
their  offset addressing  boundaries for  a given  procedure,  i.e. it
returns  $(p,  k,  a_1,  a_2,  \ldots,  a_k,  a_{k+1})$  for  a  given
procedure.
\begin{eqnarray}
  \label{eq:full:seg:pg:algo}
  &~&
  doPageSeg(P, M^\prime, M, g)\ \equiv
  \nonumber \\
  &~& \quad
  \biggl(
  \forall i \in \mathbb{N}\ \vert\ (1 \le i \le \#P)\ :\ 
  \nonumber \\
  &~& \quad \quad 
  (\mathrm{Let}\ (p_i \names Sel(Org(Enum(P, \mathbb{N})),\ i))\
  \mathrm{in} 
  \nonumber \\
  &~& \quad \quad \quad
  (\mathrm{Let}\ (segData \names getSegsData(p_i))\ \mathrm{in} 
  \nonumber \\
  &~& \quad \quad \quad \quad 
  (\mathrm{Let}\ (st \names \{segTab(M^\prime, p_i, segData)\})\
  \mathrm{in}  
  \nonumber \\
  &~& \quad \quad \quad \quad \quad 
  (\forall j \in \mathbb{N}\ \vert\ (1 \le j \le \#st)\ :
  \nonumber \\
  &~& \quad \quad \quad \quad \quad \quad
  (s_j \names Sel(Enum(st, \mathbb{N}), j))\ \wedge\ 
  \nonumber \\
  &~& \quad \quad \quad \quad \quad \quad
  (pg \names pageTab(M, s_j, g))
  ))))\biggr)
\end{eqnarray}

As     stated    above,     the     outermost    quantification     in
Eq.(\ref{eq:full:seg:pg:algo}) runs over all the procedures in the set
$P$.   Operations  like  $makeSegs$  and $makeFrame$  are  within  the
algorithms  $segTab$  and   $pageTab$.   They  preserve  the  required
invariants.  The  principle of binding permits  a different expression
of  Eq.(\ref{eq:full:seg:pg:algo})  which is  used  in practice.   The
$makeFrame$   algorithm  can   be   hoisted  out   of  the   outermost
quantification  over the procedures  in Eq.(\ref{eq:full:seg:pg:algo})
since it  is independent of  the procedures.  Given a  framed physical
memory,  the $pageTab$ algorithm  in Eq.(\ref{eq:proc:build:page:tab})
would use  an existential quantifier over  free frames to  bind with a
page.  The generic $Org$ function in Eq.(\ref{eq:basic:mgmt}) has been
replaced by the detailed functions that organize the virtual memory in
segments   and   physical   memory    in   frames.    The   $Org$   in
Eq.(\ref{eq:full:seg:pg:algo}) is  over the procedures in  $P$, and is
defined by the scheduler.

\end{document}